\documentclass[%
 aip,
 amsmath,amssymb,
reprint, onecolumn %%%%% Comment onecolumn to override one column format %%%%%%%
]{revtex4-1}

\usepackage{graphicx}% Include figure files
\usepackage{dcolumn}% Align table columns on decimal point
\usepackage{bm}% bold math
\usepackage[utf8]{inputenc}
\usepackage[T1]{fontenc}
\usepackage{mathptmx}
\usepackage{etoolbox}

\usepackage{footnotebackref}

\makeatletter
\def\@email#1#2{%
 \endgroup
 \patchcmd{\titleblock@produce}
  {\frontmatter@RRAPformat}
  {\frontmatter@RRAPformat{\produce@RRAP{*#1\href{mailto:#2}{#2}}}\frontmatter@RRAPformat}
  {}{}
}%
\makeatother
\begin{document}

\preprint{AIP/123-QED}

\title[]{Discovering Novel Halide Perovskite Alloys using Multi-Fidelity Machine Learning and Genetic Algorithm}
% Force line breaks with \\

\author{Jiaqi Yang}
%\email{yang1494@purdue.edu}
\affiliation{Materials Engineering, Purdue University, 701 W Stadium Ave, West Lafayette, 47907, Indiana, USA}

\author{Panayotis Manganaris}
%\email{pmangana@purdue.edu}
\affiliation{Materials Engineering, Purdue University, 701 W Stadium Ave, West Lafayette, 47907, Indiana, USA}

\author{Arun Mannodi-Kanakkithodi}
%\homepage{http://www.Second.institution.edu/~Charlie.Author.}
\affiliation{Materials Engineering, Purdue University, 701 W Stadium Ave, West Lafayette, 47907, Indiana, USA}
\email{amannodi@purdue.edu}

\footnote{J.Y. and P.M. contributed equally to this work}

\date{\today}% It is always \today, today,

\begin{abstract}
Expanding the pool of stable halide perovskites with attractive optoelectronic properties is crucial to addressing current limitations in their performance as photovoltaic (PV) absorbers. In this article, we demonstrate how a high-throughput density functional theory (DFT) dataset of halide perovskite alloys can be used to train accurate surrogate models for property prediction and subsequently perform inverse design using genetic algorithm (GA). Our dataset consists of decomposition energies, band gaps, and photovoltaic efficiencies of nearly 800 pure and mixed composition ABX$_3$ compounds from both the GGA-PBE and HSE06 functionals, and are combined with $\sim$ 100 experimental data points collected from the literature. Multi-fidelity random forest regression models are trained on the DFT + experimental dataset for each property using descriptors that one-hot encode composition, phase, and fidelity, and additionally include well-known elemental or molecular properties of species at the A, B, and X sites. Rigorously optimized models are deployed for experiment-level prediction over > 150,000 hypothetical compounds, leading to thousands of promising materials with low decomposition energy, band gap between 1 and 2 eV, and efficiency > 15\%. Surrogate models are further combined with GA using an objective function to maintain chemical feasibility, minimize decomposition energy, maximize PV efficiency, and keep band gap between 1 and 2 eV; hundreds more optimal compositions and phases are thus discovered. We present an analysis of the screened and inverse-designed materials, visualize ternary phase diagrams generated for many systems of interest using ML predictions, and suggest strategies for further improvement and expansion in the future. 
\end{abstract}

\maketitle

\section*{INTRODUCTION}

The importance of engineering and optimizing halide perovskites (HaPs) for use as absorbers in next-generation solar cells cannot be overstated. There is a deluge of experimental and computational work on this topic emerging every single day \cite{HaP1,HaP2,HaP3,HaP4,HaP5}; even if such attempts appear to be plateauing, incremental improvements are important and motivate more comprehensive investigations. Record power conversion efficiencies were recently reported by multiple sources for perovskite-based tandem solar cells \cite{perovs_pv1,perovs_pv2}. Curiously, there seems to be ever more room for better performance by tailoring the ABX$_3$ composition (for canonical 3D crystalline perovskites) in terms of the number and nature of species that inhabit the A/B/X sites, dopants, phase stability, interfaces, and defects \cite{MRS_Bulletin_JY_Arun}. Of course, HaPs may also adopt the double perovskite structure or a 2D phase via the use of organic spacers, and be used in both single-junction or tandem solar cells with wide or low band gap semiconductors. \\

Our research group has made several contributions to the HaP literature over the last few years, reporting updates on discovering promising new materials using density functional theory (DFT)-based computational screening and machine learning (ML) \cite{MRS_Bulletin_JY_Arun,Mannodi_HaP_1,Mannodi_HaP_2,Mannodi_HaP_3}. In 2022, we published a major study on using DFT-ML to design novel B-site mixed ABX$_3$ HaPs with desired stability, band gap, photovoltaic (PV) figure of merit, and defect tolerance \cite{Mannodi_HaP_1}. In a review paper in 2022, we covered several similar efforts from the literature utilizing high-throughput (HT) computations and ML models trained on small, medium, or large datasets of perovskite properties, to drive the discovery and understanding of HaPs \cite{MRS_Bulletin_JY_Arun}. Recently, we published another comprehensive study on generating one of the largest known DFT datasets of pseudo-cubic HaP alloys, with mixing allowed at A, B, or X sites, from multiple DFT functionals, resulting in (a) a thorough analysis of how common cation and anion choices and types of mixing affect the stability and optoelectronic properties, (b) an understanding of how different levels of theory in DFT reproduce experimentally measured properties, and (c) an open-access dataset that can be utilized by anybody for data mining and ML endeavors \cite{Mannodi_HaP_2}. We followed up this study by extending the DFT dataset to a series of non-cubic HaPs in multiple prototype phases, using multiple semi-local and non-local DFT functionals, as well as applying varying degrees of octahedral distortions and strain and different types of ionic ordering in alloys \cite{Mannodi_HaP_3}. \\

\begin{figure}[ht]
\includegraphics[width=1.0\linewidth]{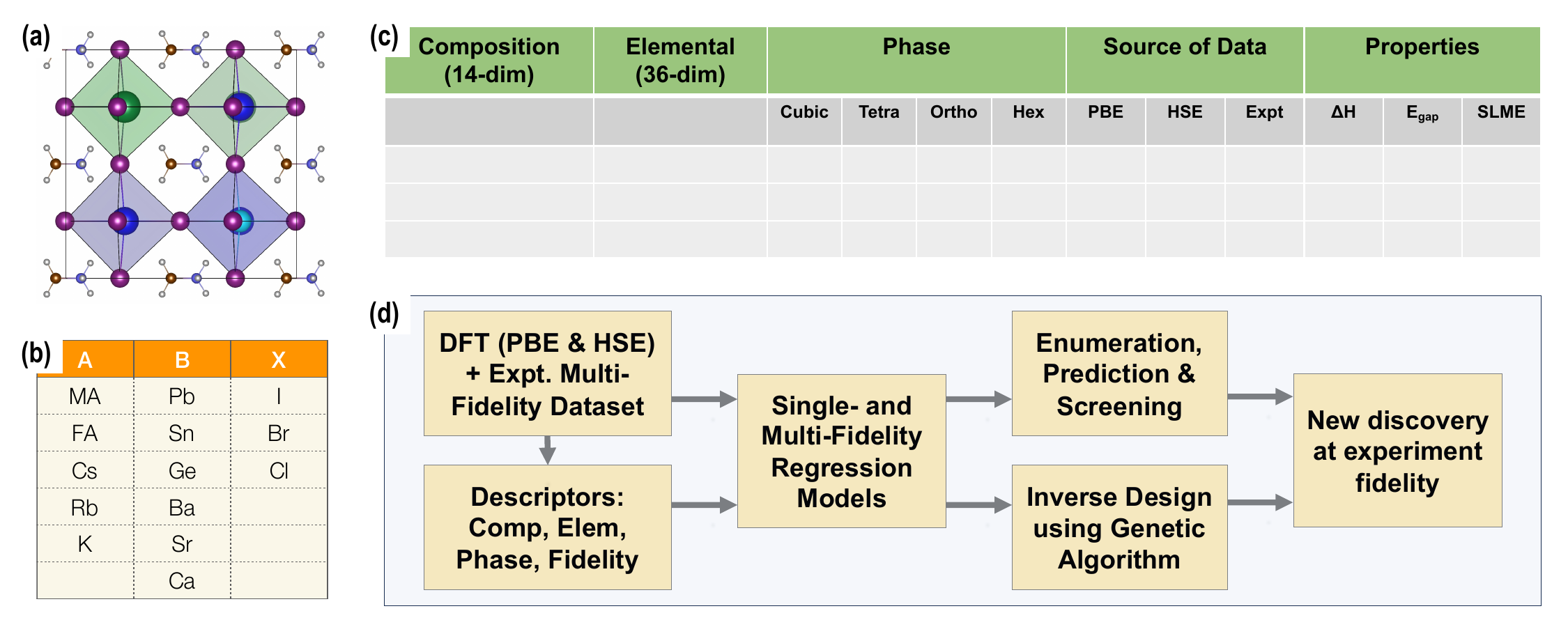}
\caption{\label{fig:outline} (a) An example pseudo-cubic 2x2x2 HaP supercell. (b) The chemical space of ABX$_3$ HaPs studied in this work. (c) The perovskite dataset formatted for multi-fidelity ML, including all inputs (composition, elemental properties, phase, and fidelity) and outputs (decomposition energy, band gap, and PV efficiency or SLME). (d) The perovskite design workflow showing data formatting, regression optimization, prediction, GA, and screening steps.}
\end{figure}

In the present contribution, we build upon our prior work and present multi-fidelity ML regression models \cite{mfml} trained on our existing multi-phase multi-functional HaP dataset of $\sim$ 10$^3$ points or so, leading to accurate predictions across hundreds of thousands of possible compounds, screening of promising candidates, and inverse design using genetic algorithm (GA) to expand the scope of materials selection beyond HT-screening. We create a fusion dataset of HaPs containing three properties, namely the decomposition energy, electronic band gap, and spectroscopic limited maximum efficiency (SLME) \cite{slme_1}, estimated from two types of DFT functionals---the semi-local GGA-PBE (570 data points), and non-local hybrid HSE06 with spin-orbit coupling (SOC) (347 data points). This dataset of 917 DFT points is further enhanced with 97 experimental data points collected from the literature \cite{exp_almora,exp_jacobsson}, reporting HaP compositions and their measured band gap and photo-conversion efficiency (PCE). \textbf{Figure \ref{fig:outline}(a)} shows a standard pseudo-cubic 2$\times$2$\times$2 perovskite supercell, and \textbf{Figure \ref{fig:outline}(b)} shows the chemical space considered in our work in terms of A, B, and X species, with MA and FA representing organic molecules methylammonium and formamidinium, respectively. \textbf{Figure \ref{fig:outline}(c)} further shows the format of the dataset, with every compound represented in terms of a 14-dimensional composition vector (capturing fractions of the 5 A species, 6 B species, and 3 X species in any compound), a 36-dimensional ``elemental properties" vector \cite{Mannodi_HaP_1,Mannodi_HaP_2} (12 weight-averaged properties each for A, B, and X site species, such as ionic radius, electron affinity, etc.), and one of four prototype perovskite phases it could adopt---cubic, tetragonal, orthorhombic, or hexagonal. Additional columns represent the source of the data (PBE, HSE, or experiment), and the properties of interest: decomposition energy or $\Delta$H, band gap or E$_{gap}$, and SLME (or PCE from experiment). \\

The use of ``multi-fidelity" learning \cite{mfml,mf1,mf2} (or multi-task learning \cite{mfml2,mfml3}) is motivated by the fact that different DFT functionals work well for different HaP compositions, and accuracy compared with experiments is not as trivial as one would imagine. While GGA-PBE, including variants such as PBEsol (improved PBE for solids \cite{PBEsol}) and PBE-D3 (explicit van der Waals corrections \cite{PBE-D3}), reproduces lattice parameters and stability reasonably well, it is typically inaccurate for electronic, optical, and defect properties \cite{MRS_Bulletin_JY_Arun}. Hybrid HSE06 is much better for optoelectronic properties, but is expensive, especially when SOC is incorporated for the relativistic effects of heavy atoms such as Pb \cite{Mannodi_HaP_1,Mannodi_HaP_2}. For hybrid organic-inorganic perovskites (HOIPs) such as MAPbI$_3$ and MAPbBr$_3$, PBE without SOC often reproduces the experimental band gap as well as HSE+SOC. While HSE+SOC should work for inorganic HaPs in general, this is not always the case, as the mixing fraction $\alpha$ in HSE06 (default value of $\alpha$ = 0.25) could itself be tuned; e.g., it was shown that for CsPbI$_3$, $\alpha$ = 0.41 reproduces the band gap more accurately \cite{CsPbI3_hse}, while $\alpha$ = 0.50 works best for cubic FAPbI$_3$ \cite{Mannodi_HaP_3}. We posit that the true relationship between the perovskite chemistry, DFT functional, and experimental properties, is enormously complex, and by learning from a fusion dataset containing properties from multiple levels of theory as well as from experiments, comprehensive experiment-fidelity predictions could be achieved across a wide chemical space. Such models would exploit inherent correlations between DFT and experiments, as well as expand the reach of experimental predictions beyond the range of chemistries for which measured data is currently available, provided DFT data is indeed available in these unexplored spaces. \\

The remainder of this manuscript presents details of the ML approaches and subsequent screening and inverse design, followed by a systematic discussion of the results. The overall methodology is shown in \textbf{Figure \ref{fig:outline}(d)}, going from compiling the DFT+Expt dataset of properties and descriptors to training several single- and multi-fidelity regression models, resulting in enumeration, prediction and screening of promising candidates and inverse design of new compounds using GA, closing the experiment-fidelity perovskite design loop. We discuss the accuracy and merits of all single- and multi-fidelity regression models based on random forest regression, and furthermore, present an analysis of the best compounds from screening and GA, in terms of the frequencies of occurrence of various chemical species and different types of mixing at cation or anion sites. All data and models are openly available to the community and will serve not only the discovery of novel HaPs for optoelectronic applications, but also provide a fertile playground for testing of a variety of ML techniques. \\

\begin{figure}[ht]
\includegraphics[width=1.0\linewidth]{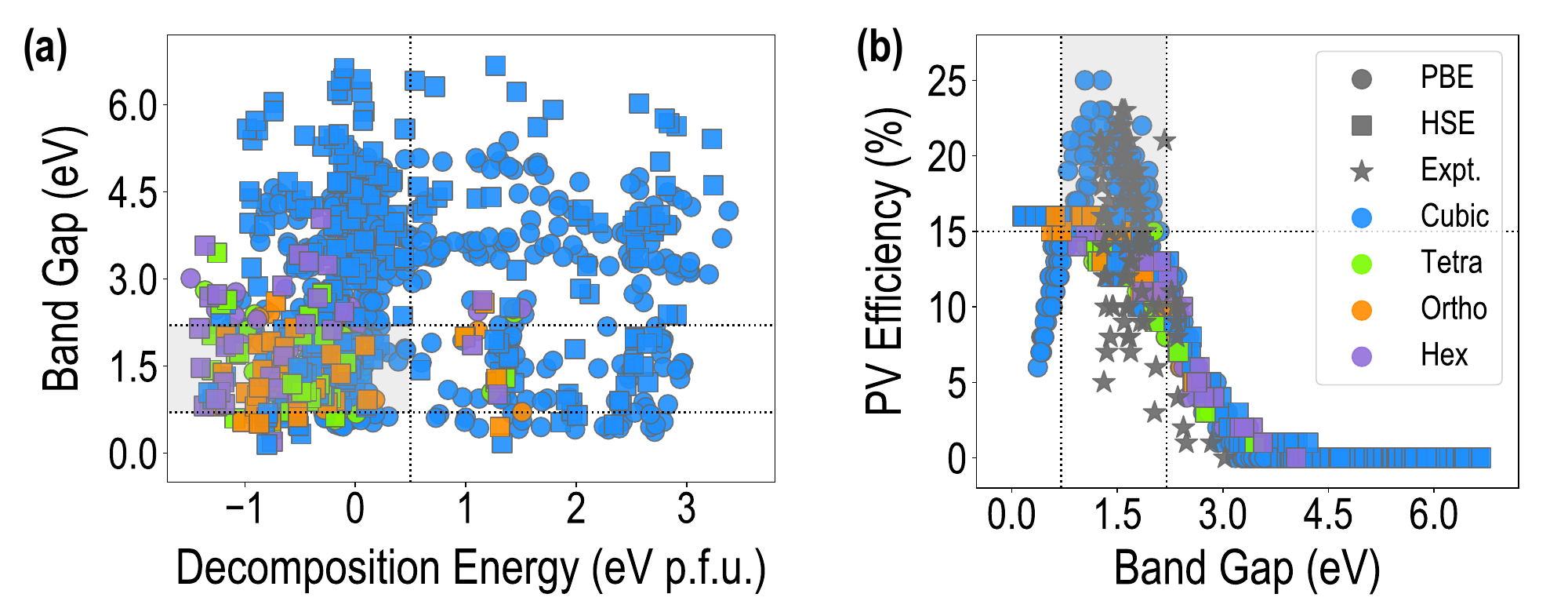}
\caption{\label{fig:dft_data} The perovskite dataset divided in terms of source of data (DFT-PBE, DFT-HSE, or experiment) and perovskite phase (cubic, tetragonal, orthorhombic, and hexagonal). (a) Band gap plotted against decomposition energy. (b) PV efficiency plotted against the band gap.}
\end{figure}

\section*{METHODS}

\subsection*{\textbf{Compiling the HaP Dataset}}

The entire HaP dataset is pictured in \textbf{Figure \ref{fig:dft_data}}, divided in terms of the source of data (PBE, HSE, or Expt) and the perovskite phase (cubic, tetragonal, orthorhombic, or hexagonal). This data is compiled from our recent publications 
\cite{Mannodi_HaP_1,Mannodi_HaP_2,Mannodi_HaP_3}. A majority of the data is for the cubic phase as a result of our initial high-throughput investigation \cite{Mannodi_HaP_1,Mannodi_HaP_2}, whereas the non-cubic data was generated in follow-up work \cite{Mannodi_HaP_3}. 570 data points are from PBE, 347 from HSE, and 97 from experiments \cite{exp_almora,exp_jacobsson}, resulting in a combined dataset of 1014 points. The decomposition energy ($\Delta$H) shows how likely the ABX$_3$ compound is to decompose to AX and BX$_2$ phases, with a mixing entropy contribution included as well, giving a convenient per formula unit (p.f.u.) metric for perovskite stability \cite{Mannodi_HaP_1,Mannodi_HaP_2}. E$_{gap}$ comes from accurate PBE and HSE electronic structure computations using dense k-point meshes, and from the experimental literature where techniques ranging from UV-vis absorption to photoluminescence spectroscopy have been applied \cite{exp_almora,exp_jacobsson}. SLME is derived at 5$\mu$m sample thickness from the DFT-computed optical absorption spectrum based on previously developed approaches \cite{slme_1,slme_2}, and is combined with measured PCE values reported from solar cells based on different HaP compositions. All DFT data are restricted to HaP compositions with any A, B, or X constituent (shown in \textbf{Figure \ref{fig:outline}(b)}) occurring only in fractions of \textit{n/8} (\textit{n} = 0, 1, 2, ... 8), such that geometry optimization and subsequent electronic and optical calculations could be performed using the special quasirandom structures (SQS) approach \cite{SQS} in 2$\times$2$\times$2 (cubic) or 2$\times$2$\times$1 (tetra, ortho, hex) supercells. Other specific DFT details and data analysis can be found in past publications \cite{Mannodi_HaP_1,Mannodi_HaP_2,Mannodi_HaP_3}. \\

The E$_{gap}$ vs $\Delta$H plot in \textbf{Figure \ref{fig:dft_data}(a)} shows that while there are dozens of compounds from all phases that lie in the $\Delta$H < 0.5 eV (a relaxed threshold) and 1 eV < E$_{gap}$ < 2 eV range, a majority of the compounds, primarily cubic, are in the undesirable ranges. \textbf{Figure \ref{fig:dft_data}(b)} shows that PV efficiencies peak around E$_{gap}$ $\sim$ 1.5 eV and the favorable region (SLME/PCE > 15\%) is dominated by cubic compounds, though that might just be a factor of the current dataset not containing as many non-cubic structures. From both plots, it is evident that only 10 to 20\% of the entire dataset of 1014 points show desired stability and optoelectronic properties. The compounds in this dataset cover the 14-dimensional ABX$_3$ chemical space adequately (including 5 A species, 6 B species, and 3 X species), in terms of how often and in what mixing fraction any particular species appears in the compound \cite{Mannodi_HaP_2,Mannodi_HaP_3}. In reality, this space is practically infinite, as mixing fractions could be as low or high as possible (and not just \textit{n/8} fractions), and any number of ions could be mixed together to create high entropy perovskites. Thus, a major motivation of this work is that learning from a modest dataset of the order of 10$^3$ points, one could, in theory, make predictions for 10$^5$ to 10$^6$ points (or beyond) which would include all intermediate compositions and mixing fractions missing from the original dataset, and perform more comprehensive screening and design. \textbf{Table \ref{table:dataset}} presents details of the PBE, HSE, and experimental datasets in terms of the number of data points for each phase and the available properties. \\

\begin{table}[t]
    \centering
%    \begin{adjustbox}{width=\textwidth}
    \begin{tabular}{|c|c|c|c|}
    \hline
        \textbf{Data Fidelity} & \textbf{Total Data Points} & \textbf{Phases (Data Points)} & \textbf{Properties} \\ \hline
        PBE  &  570  &  Cubic (469), Tetra (34), Ortho	(37), Hex (30)  &  $\Delta$H, E$_{g}$, SLME  \\
        HSE  &  347  &  Cubic (246), Tetra (34), Ortho	(37), Hex (30)  &  $\Delta$H, E$_{g}$, SLME  \\
        Expt.  &  97  &  Phases assigned from DFT-ML Predictions  &  E$_{g}$, PCE  \\ \hline
    \end{tabular}
%    \end{adjustbox}
    \caption{\label{table:dataset} Description of the perovskite dataset used for training ML models in this work, in terms of total number of data points, data points per perovskite phase, and the properties available.}
\end{table}

\subsection*{\textbf{Training Single- and Multi-Fidelity Surrogate Models}}

The dataset is formatted as shown in \textbf{Figure \ref{fig:outline}(c)} and fed into a random forest regression (RFR) algorithm for rigorous training and optimization of predictive models for each property. Standard data science and ML practices are applied: an 80-20 train-test split is used for the entire dataset, 5-fold cross-validation is applied, and grid-based hyperparameter optimization is performed. Relevant modules are imported from the Scikit-learn library \cite{scikit-learn} and all our code is available on Github \cite{my_repo}. Four types of descriptors are used as input (\textbf{X}) to the RFR models for predicting property \textbf{Y} ($\Delta$H, E$_{gap}$, SLME/PCE): 

\begin{enumerate}
    \item 14-dimensional composition vector: This encodes the HaP composition in terms of the fraction (between 0 and 1) of every A/B/X species in the compound.
    \item 36-dimensional elemental properties vector: This uses 12 distinct well-known properties (the full list is provided in Table S1 in the SI) to represent the A, B, and X site constituents, using a weighted fraction when there is mixing at any site.
    \item 4-dimensional phase vector: This one-hot encodes whether the perovskite phase is cubic, tetragonal, orthorhombic, or hexagonal.
    \item 0-dim, 2-dim, or 3-dim fidelity vector: This one-hot encodes the source of the data (PBE, HSE, or Expt) \cite{mfml,mfml2}. \\
\end{enumerate}

Three different types of RFR models are trained:

\begin{enumerate}

    \item PBE, HSE, and Expt. single-fidelity (SF) models: The source of the data is not an input here, so a total of 54 dimensions are used as X and separate models are trained to predict $\Delta$H, E$_{gap}$ and SLME each from PBE and HSE, as well as E$_{gap}$ and PCE just from experiments. These models are subsequently referred to as PBE-sf, HSE-sf, and Expt-sf, respectively.
    
    \item PBE+HSE multi-fidelity (MF) models: A 2-dim data source vector is included to yield a 56-dimensional X, used as input to predict the three properties simultaneously from both functionals based on the dataset of 917 DFT points. The advantage here is that the reach and prediction accuracy of the smaller and more expensive HSE dataset can be enhanced by utilizing its correlations with the supposedly inferior but larger PBE dataset. These predictions will be referred to as PBE-mf1 and HSE-mf1, respectively.
    
    \item PBE+HSE+Expt MF models: A 3-dim data source vector is included to yield a 57-dimensional X, which is used as input to predict E$_{gap}$ and SLME or PCE simultaneously from PBE, HSE, and Expt based on the combined DFT-Expt dataset of 1014 points. A clear advantage is that the applicability of the much smaller experimental dataset can be enhanced by utilizing its correlations with the PBE and HSE data. These predictions will be referred to as PBE-mf2, HSE-mf2, and Expt-mf2, respectively. \\
    
\end{enumerate}

RFR results after training and optimization are visualized in terms of parity plots between ground truth values (actual PBE, HSE, or Expt) on the x-axis and ML predictions on the y-axis, separately for each property and each source of data. It should be noted that the ultimate test of the model's predictive power is its performance on unseen data points, and ideally, one must look at the quality of prediction for every data point when it is considered as part of the test set. To achieve this, we adopt the following ensemble-based strategy: every SF and MF RFR model is trained 5000 times across each dataset, with an 80-20 split, such that every single data point is considered in the test set approximately 1000 times. Effective test set predictions are then achieved for all points as a mean over their $\sim$ 1000 test predictions (also yielding the standard deviation, which serves as a rough uncertainty in prediction for any point). Thus, final parity plots, which will be presented and discussed later, show only test predictions for all PBE, HSE, and Expt points.

\subsection*{\textbf{Enumeration, Prediction, and Screening}}

Since only a small subset of possible ABX$_3$ compositions is used for the DFT+ Expt dataset, we generate a much larger dataset of ``hypothetical" HaP compositions by populating the 14-dim composition vector with fractions of n/8, such that the A components sum up to 1, B components sum up to 1, and X components sum up to 3. A similar approach was applied in other works as well \cite{Mannodi_HaP_1,Mannodi_HaP_4}. To keep this combinatorial dataset somewhat tractable, we restrict it to only one type of mixing at a time (i.e., there will not be mixing at A site and B or X sites simultaneously) and only mixing fractions of \textit{n/8}. This leads to a total of 37,785 unique compositions including thousands of A-site mixed, B-site mixed, and X-site mixed compounds each, which is transformed to 151,140 data points considering them in four phases each. All compounds are converted into descriptors and fed to the best RFR models, eventually resulting in prediction of their PBE/HSE-fidelity $\Delta$H as well as experiment-fidelity E$_{gap}$ and PV efficiency. Screening is performed to obtain the subset of these > 150,000 compounds with low decomposition energy, band gap between 1 and 2 eV, and PV efficiency > 15\%. \\

\subsection*{\textbf{Inverse Design using GA}}

Finally, we move beyond the constraints of the restricted chemical space considered for enumeration, to perform a more efficient design of HaP compositions in any phase that satisfy conditions of DFT stability and experiment-level optoelectronic properties. A constrained or clamped Genetic Algorithm (GA) is used for this purpose \cite{GA1,GA2,GA3}, using a procedure wherein hundreds of arbitrary ABX$_3$ compositions are generated, with any kind of mixing allowed at A, B, or X sites, in any of the four phases, and a complex objective function is defined for each compound taking the following factors into account:

\begin{enumerate}
    \item Chemical feasibility: Constituent fractions at A, B, and X sites should respectively sum up to 1, 1, and 3.
    \item Suitability of modeling via DFT: Mixing fractions are restricted to be n/8, n/27, or n/64 (where is a positive integer), such that any composition could be simulated using a 2$\times$2$\times$2, 3$\times$3$\times$3, or 4$\times$4$\times$4 supercell.
    \item Stability: $\Delta$H should be minimized to maintain maximum likelihood of perovskite formability. The HSE-level $\Delta$H prediction is used here.
    \item Optoelectronic properties: E$_{gap}$ should be in between 1 and 2 eV and PV efficiency should be as high as possible, with both predictions at experiment-fidelity. 
\end{enumerate}

During any GA run, new HaP compositions are generated using concepts such as crossover, elitism, and mutation, so as to minimize the objective function, yielding a plot between GA generation and objective function, eventually resulting in the best material that satisfies all criteria. Running this several hundred times leads to a massive list of attractive materials that further improve on the compounds derived from enumeration $\rightarrow$ prediction $\rightarrow$ screening. Here, we run GA multiple times with different types of inputs, such as restricting the phase to be cubic or tetragonal, restricting the B-site to consider only Pb, Sn, and Ge, or to not consider Pb at all in a bid to achieve Pb-free perovskites, etc. Final results can be visualized for different families of HaPs in terms of PV efficiency vs band gap plots for hundreds of stable compounds. The "geneticalgorithm2" package is used for running the GA models \cite{GA4}. \\

\section*{RESULTS AND DISCUSSION}

\subsection*{\textbf{DFT-Only Single-Fidelity and Multi-Fidelity RFR Models}}

\begin{figure}[ht]
\includegraphics[width=0.8\linewidth]{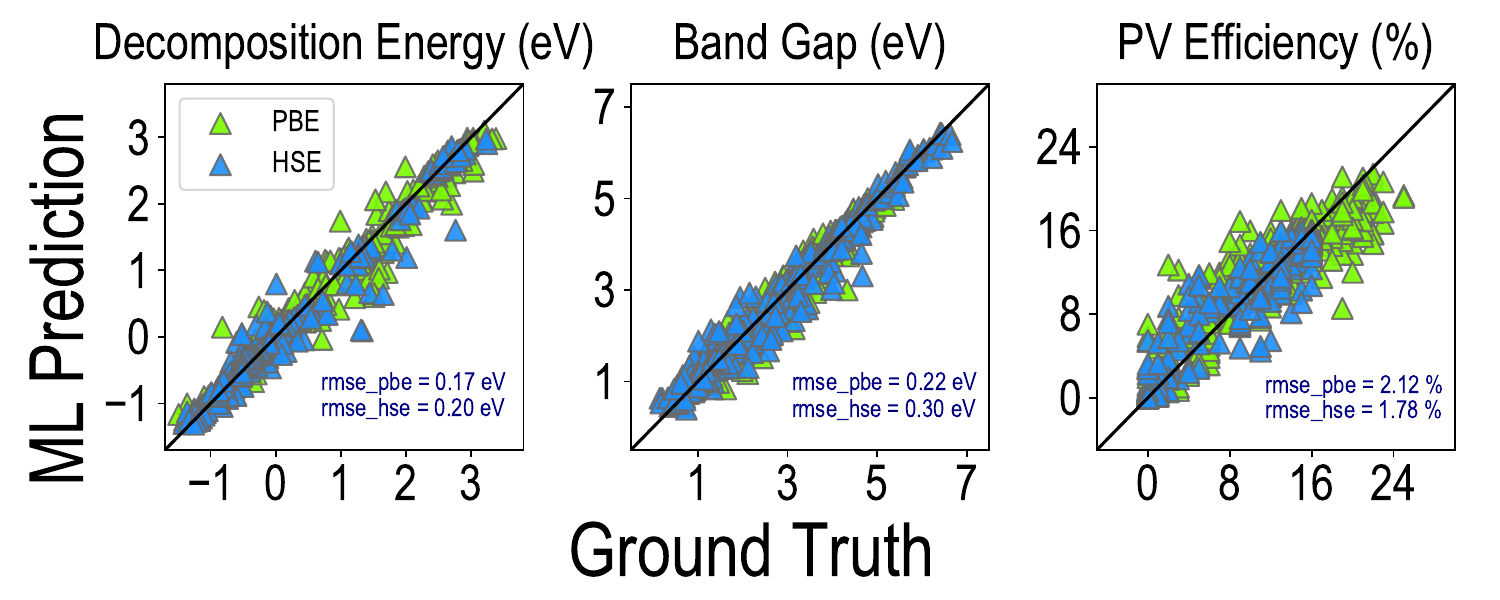}
\caption{\label{fig:ml_sf} Single-fidelity random forest regression models trained for decomposition energy, band gap, and PV efficiency, separately on the DFT-PBE and DFT-HSE datasets. Parity plots capture effective averaged test predictions (over 5000 models) for every data point, and RMSE values are shown separately for PBE and HSE.}
\end{figure}

\textbf{Figure \ref{fig:ml_sf}} shows the PBE-sf and HSE-sf models for $\Delta$H, E$_{gap}$, and PV efficiency, with PBE and HSE data points presented on the same plot. The parity plots show effective test predictions, averaged over an ensemble of 5000 models, for all 917 data points from PBE and HSE. High accuracy is achieved for both $\Delta$H and E$_{gap}$, with RMSE values of 0.17 eV and 0.23 eV respectively for PBE, and 0.20 eV and 0.30 eV respectively for HSE. These are errors of 5\% or less considering the range of values of both properties across the PBE and HSE datasets, indicating a 95\% accuracy. Given that there are between 5 and 12 atoms p.f.u. in any HaP compound, the $\Delta$H RMSE converts to $\sim$ 20-40 meV/atom, which are highly competitive with other ML formation energy predictions in the literature \cite{form-ml1,form-ml2}. The same can be said for E$_{gap}$ prediction errors between 0.2 and 0.3 eV as well \cite{gap-ml1,gap-ml2}. Predictions errors for PBE and HSE SLME are higher on average, with RMSE values of 2.12\% and 1.78\% respectively; these errors are $\sim$ 10\% of the total range of values, and show that the composition-based descriptors may not be sufficient for highly accurate prediction of PV efficiency. This is not a surprise, as the optical absorption spectra are likely sensitive to the perovskite structure and would thus affect the SLME prediction. Furthermore---especially when experimental data will be included in training---there are many factors beyond the perovskite composition alone, including the solar cell architecture, interfaces, and external conditions, that influence the PV efficiency. Nevertheless, the focus of current work is on determining upper limit efficiencies as a function of HaP compositions and eventually screening materials which are likely to show high efficiencies. \\

\begin{figure}[ht]
\includegraphics[width=0.8\linewidth]{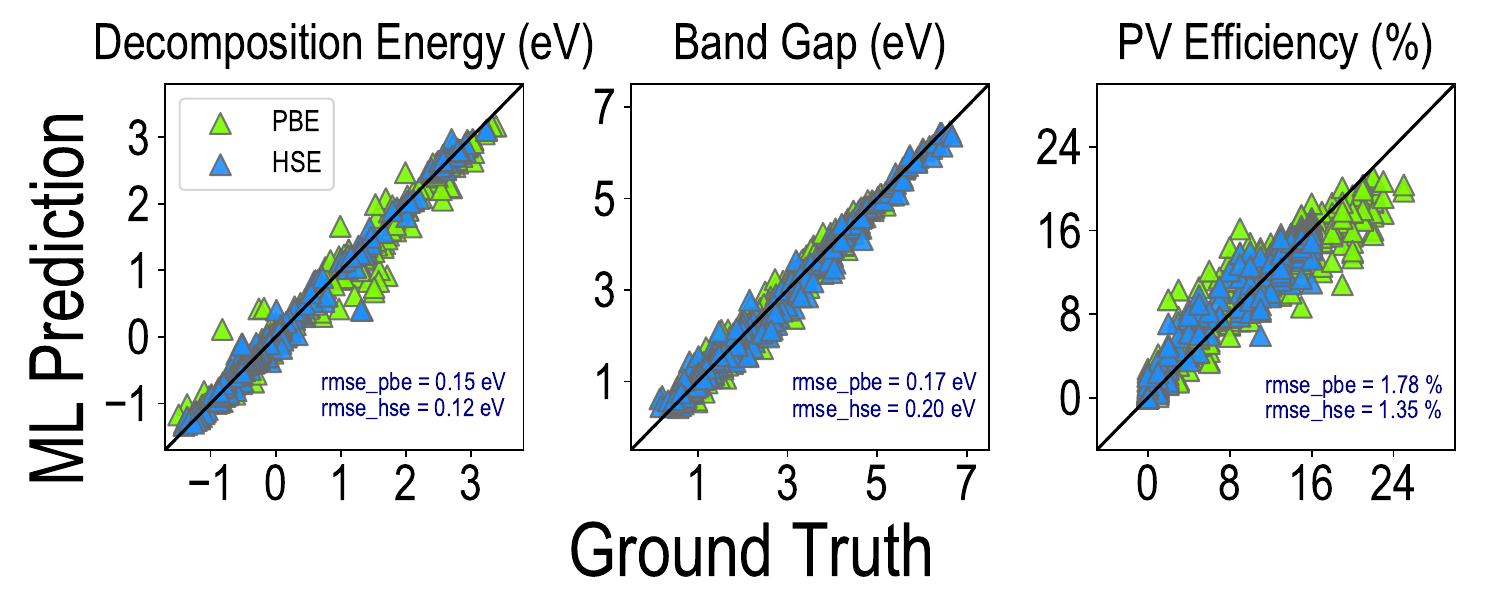}
\caption{\label{fig:ml_mf1} Multi-fidelity random forest regression models trained for decomposition energy, band gap, and PV efficiency, based on a combined DFT-PBE + DFT-HSE dataset. Parity plots capture effective averaged test predictions (over 5000 models) for every data point, and RMSE values are shown for PBE and HSE.}
\end{figure}

Next, the DFT-only MF models are trained for each property by adding 2 descriptors indicating whether the data source is PBE or HSE, leading to PBE-mf1 and HSE-mf1 predictions. Results are presented in \textbf{Figure \ref{fig:ml_mf1}}, showing an improvement in performance over the SF models, with models once again picturing averaged test-only predictions over 5000 models. PBE and HSE $\Delta$H models show RMSE values of 0.15 eV and 0.12 eV respectively. E$_{gap}$ predictions also represent slight improvement, with PBE RMSE of 0.17 eV and HSE RMSE of 0.20 eV, while SLME predictions show marked improvement with RMSE of 1.78\% for PBE and 1.35\% for HSE. While $\Delta$H and E$_{gap}$ values in the HSE dataset cover a wide range of values, it can be seen from \textbf{Figures \ref{fig:dft_data}(b)}, \textbf{\ref{fig:ml_sf}}, and \textbf{\ref{fig:ml_mf1}} that the HSE SLME values clearly occur in a smaller range than the PBE SLME values, which is an artifact of the nature of this dataset and has been explained in prior work \cite{Mannodi_HaP_2}. As such, we expect the MF models to theoretically expand the reach of the HSE predictions to new regions in the chemical space that might show higher PV efficiencies, something that may not be possible with SF models given the interpolative nature of regression models. Generally, the PBE and HSE $\Delta$H values, both SF and MF1 predictions, have a good correlation and can be equally trusted, but given the low test prediction RMSE of 0.12 eV, the HSE-mf1 $\Delta$H model should provide the best way for new prediction and screening of the bulk stability of HaPs. \\

\subsection*{\textbf{RFR Models for Experiment-Fidelity Predictions}}

Finally, multiple approaches are used for making predictions on the experimental data points, and to eventually achieve accurate and general experiment-fidelity predictions. An important shortcoming of the dataset had to be overcome here: while we could collect experimental data on nearly 100 points from the literature which included their chemical composition, band gap, and PV efficiency (and some other solar cell-relevant quantities \cite{exp_almora}), there was almost never sufficient information available about the corresponding perovskite phase. An exhaustive search of every composition in the experimental literature to determine its preferred phase is beyond this scope of the current study, though it can and should be performed in the future. Furthermore, many studies do not explicitly report the likely phase of the material, though diffraction patterns and spectroscopic measurements are reported, making it difficult to definitively assign phase information to any compound, which makes a computational estimate a better approach. As such, we apply the following strategy to create a 4-dimensional phase vector (corresponding to cubic, tetragonal, orthorhombic, or hexagonal phase) for every experimental data point: 

\begin{enumerate}
    \item The PBE-sf $\Delta$H is predicted for all 100 compounds considered in all 4 phases.
    \item Predictions are made over an ensemble of 5000 models as described earlier, and the most stable phase (which of the four phases has the lowest predicted PBE $\Delta$H?) is noted each time.
    \item A score between 0 and 1 is assigned to each of the 4 dimensions of the phase vector based on the relative frequency of occurrence of any phase as the most stable. For instance, for the compound FASnI$_3$, the cubic phase fraction is found to be 0.57 and the hexagonal phase fraction is 0.41.
    \item With all 54 dimensions (including composition, phase, and elemental property vectors) now available for all experimental data points, single-fidelity models could be trained, and this data can be added to the DFT data along with a 3-dimensional fidelity vector to train the new MF models.
\end{enumerate}

\begin{figure}[ht]
\includegraphics[width=0.8\linewidth]{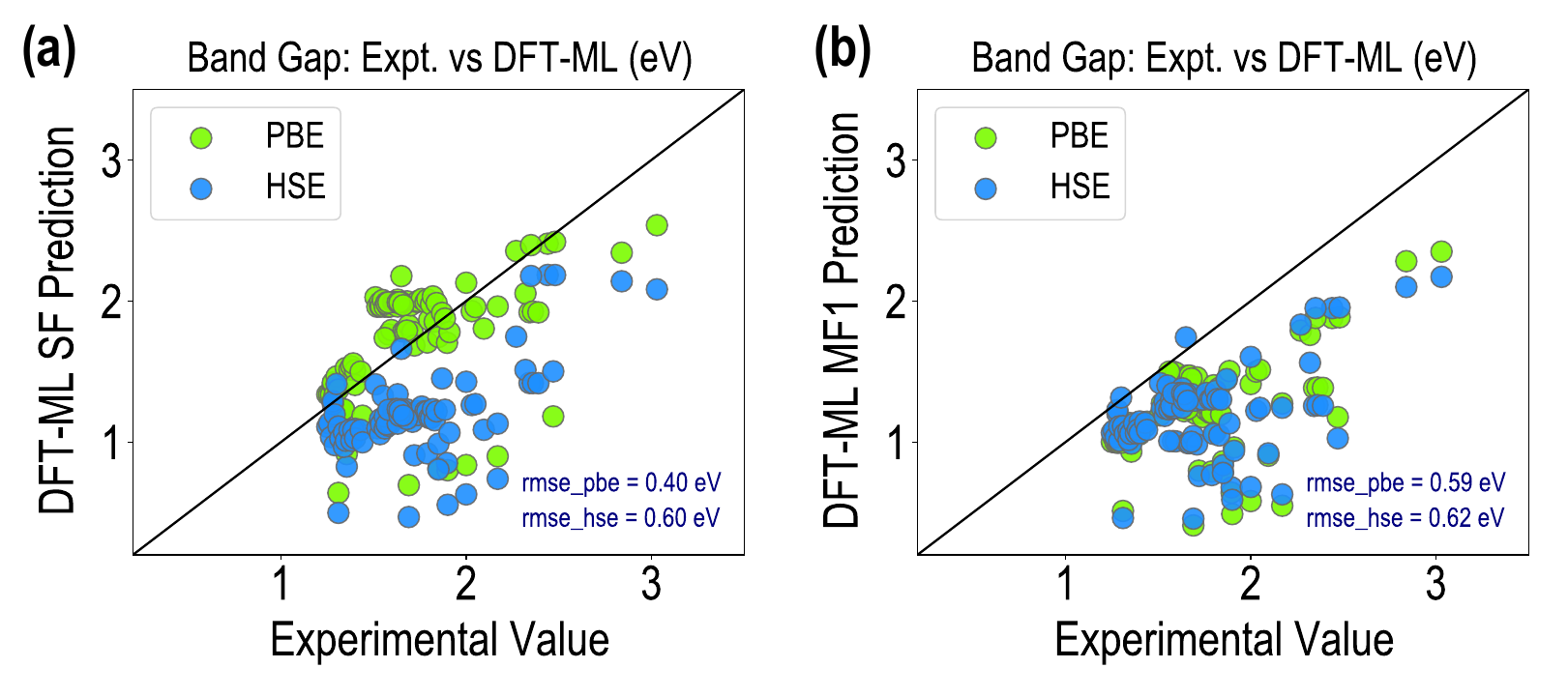}
\caption{\label{fig:expt_vs_dft} DFT-ML predictions compared to experimental values for (a) single-fidelity PBE and HSE models, and (b) DFT-only multi-fidelity models. Expt vs DFT RMSE values are shown separately for PBE and HSE.}
\end{figure}

\begin{figure}[ht]
\includegraphics[width=0.7\linewidth]{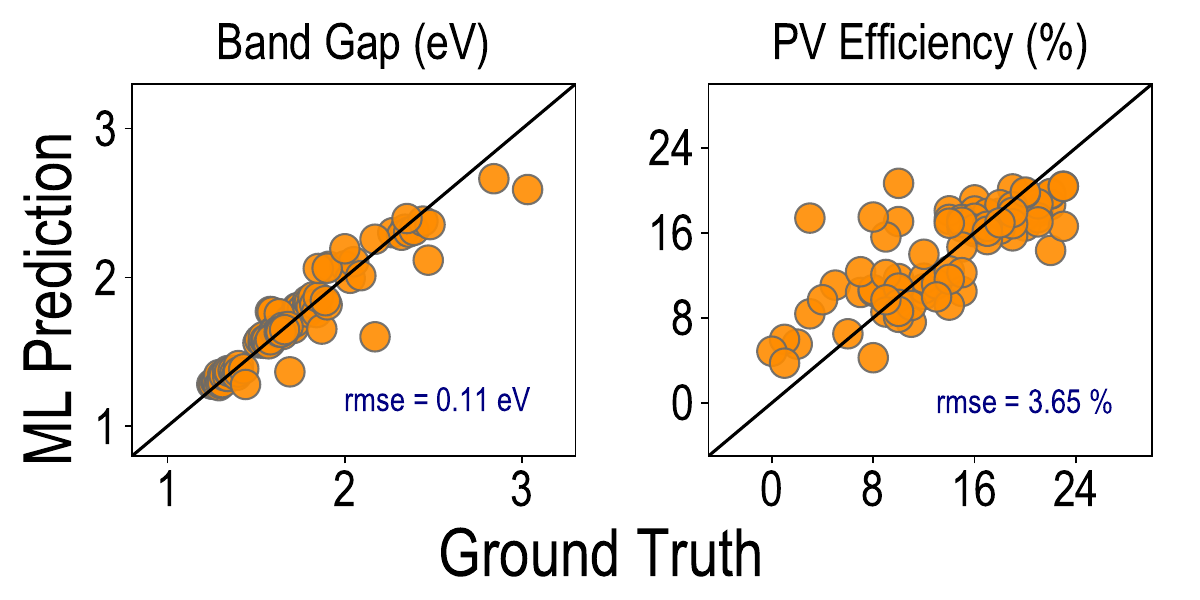}
\caption{\label{fig:expt_sf} Experiment-only single-fidelity RFR predictions for (a) band gap, and (b) PV efficiency, with effective test predictions and corresponding RMSE shown.}
\end{figure}

In the absence of any experimental data, one might wonder how the DFT-predicted E$_{gap}$ and SLME compare with measured values. A comparison of DFT vs experiments has been performed for a small set of compounds in past work, and the general understanding is that PBE and HSE errors are quite large for a lot of compounds \cite{Mannodi_HaP_1,Mannodi_HaP_2,Mannodi_HaP_3}. To examine this effect further, we used the PBE-sf, HSE-sf, PBE-mf1, and HSE-mf1 models to make predictions on all 97 experimental data points, using the 54-dimensional inputs described above. The DFT-ML SF and MF predictions of E$_{gap}$ are plotted against actual experimental values in \textbf{Figure \ref{fig:expt_vs_dft}(a)} and \textbf{(b)} respectively: it can be seen that PBE-sf predictions have the lowest RMSE of 0.4 eV, with all other predictions lying in the $\sim$ 0.6 eV RMSE range. These errors are clearly too high, and certain compounds are especially poorly predicted, with errors larger than 1.5 eV at times. Corresponding parity plots comparing PBE-sf, HSE-sf, PBE-mf1, and HSE-mf1 predictions of SLME with experimental PCE are presented in \textbf{Figure S1}, showing that there is essential no correlation between the DFT-ML and experimental values of PV efficiency. \\

Next, SF models were trained for the experimental E$_{gap}$ and PCE, and effective test predictions for all 97 points based on an ensemble of 5000 models are presented in \textbf{Figure \ref{fig:expt_sf}}. E$_{gap}$ predictions show a very low RMSE of 0.11 eV, though a caveat is the lower range of values in the experimental dataset, whereas RMSE on PCE predictions is much higher at 3.72\%. Although the Expt-sf models could already be used for making experiment-fidelity predictions on hundreds of thousands of hypothetical compounds, it would be limited and erroneous given the narrow chemical space in the experimental dataset compared to the DFT dataset---where explicit care was taken to involve all chemical species in roughly equal proportions across all the compounds. Thus, a DFT-Expt MF model might serve high-throughput predictions better. \\

\begin{figure}[ht]
\includegraphics[width=0.8\linewidth]{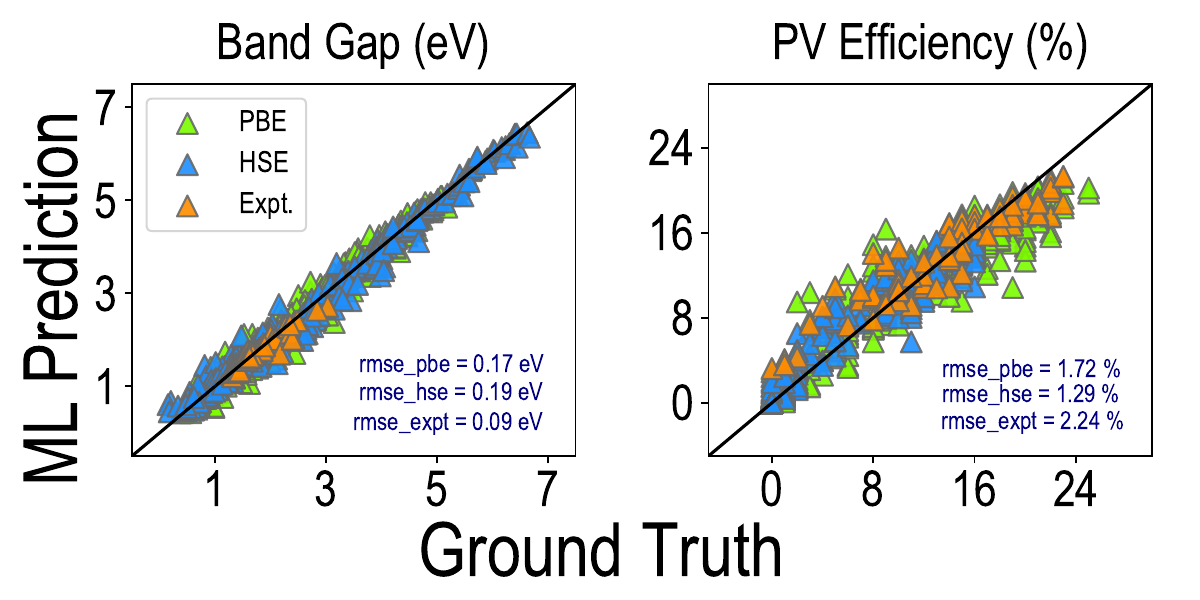}
\caption{\label{fig:ml_mf2} Multi-fidelity random forest regression models trained for the band gap and PV efficiency from DFT-PBE, DFT-HSE, and experiment. Parity plots capture effective test predictions averaged over 5000 models, and RMSE values are shown separately for PBE, HSE, and Expt.}
\end{figure}

PBE-mf2, HSE-mf2, and Expt-mf2 prediction performances are pictured in \textbf{Figure \ref{fig:ml_mf2}}. PBE and HSE E$_{gap}$ predictions show similar RMSE of 0.17 eV and 0.19 eV respectively, similar to the MF1 models. The same is true for SLME predictions where the PBE RMSE is 1.72\% and the HSE RMSE is 1.29\%. Expt-mf2 E$_{gap}$ RMSE is 0.14 eV and PCE RMSE is 3.62\%. These predictions are again quite similar to the Expt-sf performances, but are vastly more useful for their applicability across new regions of the HaP chemical space. We believe that the MF2 models learn inherent relationships between Expt, PBE, and HSE values, and are thus capable of making experiment-fidelity predictions for chemistries that have been studied from PBE and/or HSE. Table \ref{table:RMSE} lists the test RMSE values for all three properties from many different RFR models, with the values in bold showing the lowest errors for any specific property at any fidelity. \\

\subsection*{\textbf{Understanding Feature Importance}}

\textbf{Figure S2} shows the Pearson coefficient of linear correlation \cite{PearsonCorr} plotted between the 56-dimensional vector representing every compound in the DFT-mf1 dataset (PBE + HSE data), against the three properties of interest. The primary observations here are very similar to past publications \cite{Mannodi_HaP_1,Mannodi_HaP_2,Mannodi_HaP_3}. Increasing the K fraction will lead to an increase in $\Delta$H and thus make the compound less stable, while the FA fraction has the exact opposite effect. $\Delta$H is highly negatively correlated with features such as the ionic radius and atomic number of the A-site cation, showing that larger cations (to a point) are best for perovskite stability. While Ca, Sr, and Ba lead to a heavy increase in E$_{gap}$, it is very negatively correlated with the electron affinity, ionization energy, and electronegativity of the B-site cations. The SLME shows the opposite trends to what is found for E$_{gap}$, with increase in the B-site electron affinity, ionization energy, and electronegativity leading to an increase in efficiency. Interestingly, we generally find low correlation between the phase/fidelity and the properties, except for the positive correlation between the cubic phase and both the $\Delta$H and the E$_{gap}$, which is likely a consequence of the dominance of cubic structures in the dataset that means a majority of the unstable compounds as well as a majority of the wide band gap compounds are cubic. \\

To understand the contributions of various features better, we plotted the feature importance values for all descriptor dimensions as obtained from the best RFR models, in \textbf{Figure S3}. It can be seen that the main contributions to $\Delta$H are from the A and X species, as well as from the elemental properties that highlight the size of the A-site cations and X-site anions---similar to the observations from \textbf{Figure S2}. For E$_{gap}$ and PV efficiency, there is an obvious dominance from the B-site electronegativity followed by the electron affinity and ionization energy, showing that the optoelectronic properties are heavily determined by the B-site cations (and a little bit from the X-site anions) and have almost no sensitivity to the A-site species. \textbf{Figure S4} shows the E$_{gap}$ and PV efficiency for the entire dataset of 1014 points plotted against the B-site electronegativity, showing a rough correlation, especially for the DFT data. \\

\begin{table}[!ht]
    \centering
    \begin{tabular}{|c|c|c|c|}
    \hline
        \textbf{RFR Model} & \textbf{$\Delta$H RMSE (eV)} & \textbf{E$_{gap}$ RMSE (eV)} & \textbf{PV Efficiency RMSE (\%)} \\ \hline
        PBE-sf & 0.17 & 0.23 & 2.12 \\ \hline
        HSE-sf & 0.20 & 0.30 & 1.78 \\ \hline
        Expt-sf & - & 0.11 & 3.65 \\ \hline
        PBE-sf vs Expt. & - & 0.40 & 5.45 \\ \hline
        HSE-sf vs Expt. & - & 0.60 & 5.48 \\ \hline
        HSE-sf vs Expt., shifted & - & 0.31 & - \\ \hline
        PBE-mf1 & \textbf{0.15} & 0.17 & 1.78 \\ \hline
        HSE-mf1 & \textbf{0.12} & 0.20 & 1.35 \\ \hline
        PBE-mf1 vs Expt. & - & 0.59 & 5.28 \\ \hline
        HSE-mf1 vs Expt. & - & 0.62 & 5.50 \\ \hline
        PBE-mf1 vs Expt., shifted & - & 0.34 & - \\ \hline
        HSE-mf1 vs Expt., shifted & - & 0.36 & - \\ \hline
        PBE-mf2 & - & \textbf{0.17} & \textbf{1.72} \\ \hline
        HSE-mf2 & - & \textbf{0.19} & \textbf{1.29} \\ \hline
        Expt-mf2 & - & \textbf{0.09} & \textbf{2.24} \\ \hline
    \end{tabular}
    \caption{\label{table:RMSE} Root mean square error (RMSE) in test set prediction for the three properties from multiple single- and multi-fidelity models.}
\end{table}

\begin{figure}[ht]
\includegraphics[width=0.7\linewidth]{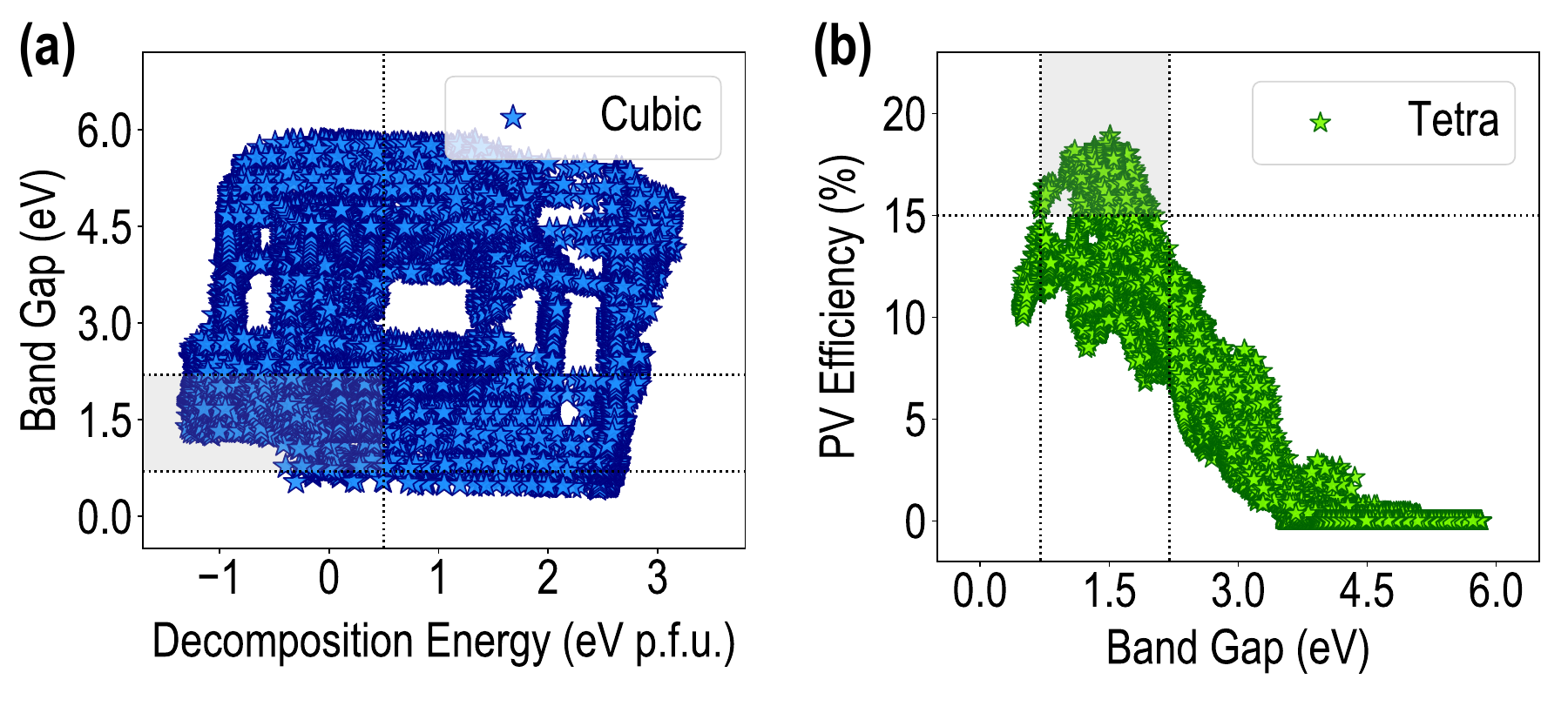}
\caption{\label{fig:out_pred} A glimpse of high-fidelity predictions made over $\sim$ 150,000 HaPs: (a) Predicted band gap (from Expt-mf2) plotted against predicted decomposition energy (from HSE-mf1) for cubic compounds, and (b) predicted PV efficiency (from Expt-mf2) plotted against predicted band gap for tetragonal phase compounds.}
\end{figure}

\begin{figure}[ht]
\includegraphics[width=0.7\linewidth]{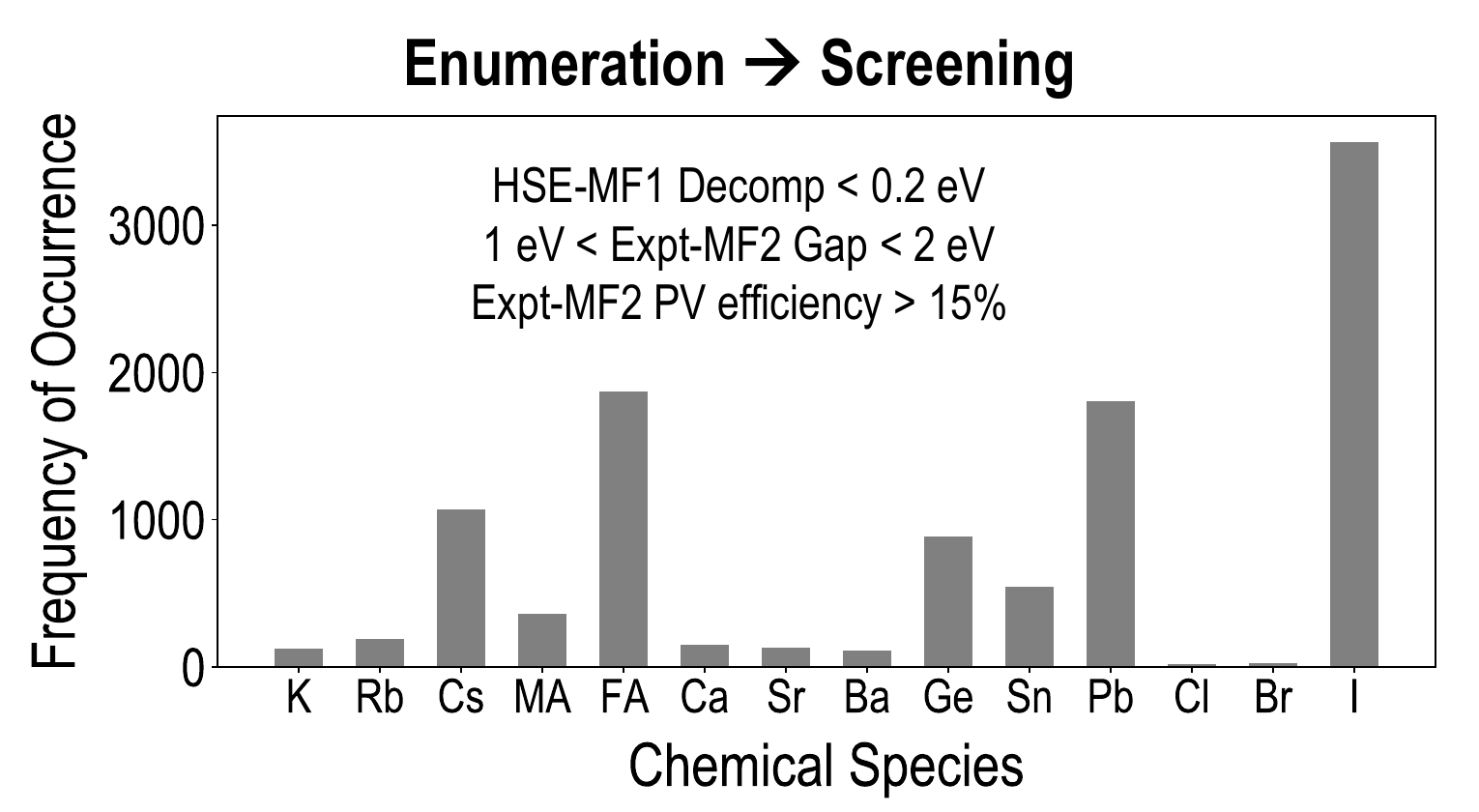}
\caption{\label{fig:screen_freq} Frequencies of occurrence of the 14 chemical species within the list of 3610 promising compounds obtained via enumeration, prediction, and screening.}
\end{figure}

\subsection*{\textbf{Prediction and Screening at Experiment-Fidelity}}

Next, predictions are made for the expanded set of 151,140 compounds populated across the same HaP chemical space. It should be noted that for any given compound, PBE or HSE computations would at least require at least 12 to 24 hours of CPU time (on approximately 128 cores) for estimating the three properties, which would correspond to over 200 years of computing time for > 150,000 compounds. Experimental investigation of all these materials would take even longer. The ML predictions, on the other hand, require only seconds for several compounds, and can be made for hundreds of thousands of compounds in mere minutes. \textbf{Figure \ref{fig:out_pred}} along with \textbf{Figures S5} and \textbf{S6} show the HSE-mf1 $\Delta$H, Expt-mf2 E$_{gap}$, and Expt-mf2 PCE predicted for the 37,785 unique compositions in four different phases, as $\Delta$H vs E$_{gap}$ and E$_{gap}$ vs PV efficiency plots. \\

It can immediately be seen that (i) general shape and distribution of the plots is the same as the DFT-Expt dataset pictured in \textbf{Figure \ref{fig:dft_data}}, and (ii) there are now hundreds of more compounds that exist in the desirable ranges of $\Delta$H < 0.2 eV, E$_{gap}$ between 1 and 2 eV, and PCE > 15\%. We find that 3610 compounds out of the 151,140 fulfil each of these criteria, which represents 2.4\% of the entire population. 1180 of these compounds are cubic, 856 tetragonal, 889 orthorhombic, and 685 hexagonal. Further, a majority of these compounds are B-site mixed (2036 in numbers) while 1364 compounds are A-site mixed, and only about 200 compounds have halogen mixing. \textbf{Figures S5} and \textbf{S6} further show that the shapes of the plots look generally similar for different phases, and there are some empty regions which will likely be covered by other intermediate compositions we have not considered in this enumeration exercise. Thus, we are able to use the multi-fidelity DFT-Expt predictions to screen > 3600 compounds with multiple desired properties. A further examination of the chemical space of the screened compounds, presented in \textbf{Figure \ref{fig:screen_freq}}, shows that FA followed by Cs and MA are the most prevalent A-site cations, Pb, Ge, and Sn are most common at the B-site, while nearly all the compounds are Iodides. This is consistent with the most typical HaP compositions used in high efficiency solar cells arising from FA-MA-Cs mixing at the A-site and a predominance of Pb and I at the other sites. \textbf{Table \ref{table:screened}} shows the total number of screened compounds divided in terms of the predicted perovskite phase and type of mixing (pure, A-mixed, B-mixed, or X-mixed). \\

\subsection*{\textbf{Discovery at Experiment-Fidelity using GA}}

\begin{figure}[ht]
\includegraphics[width=0.8\linewidth]{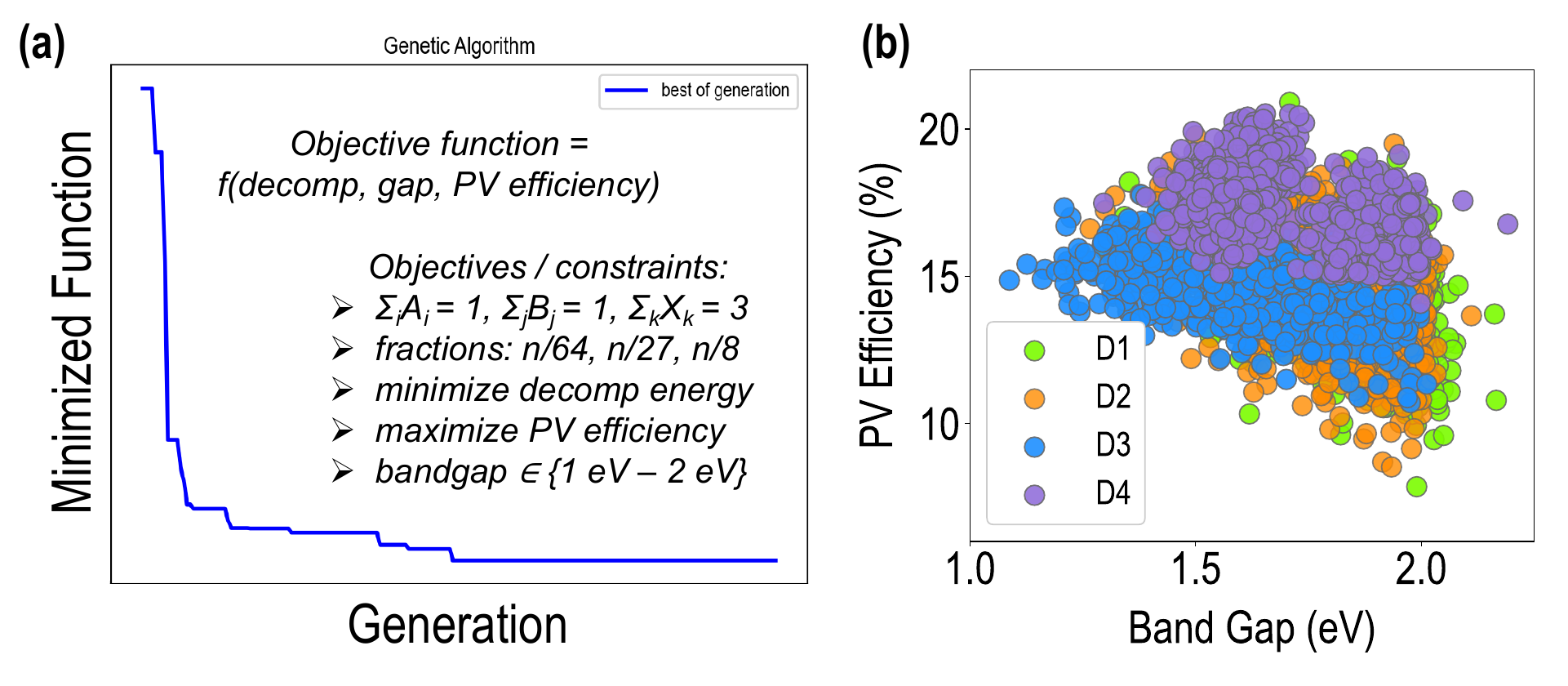}
\caption{\label{fig:GA} (a) An example GA run presented in terms of the objective function plotted against the number of generations, showing the property and chemical feasibility constraints. (b) Predicted (experiment-fidelity) PV efficiency plotted against band gap for $\sim$ 3000 compounds from several GA runs; all compositions are chemically meaningful and have decomposition energy < 0.2 eV p.f.u.}
\end{figure}

The GA results are presented in \textbf{Figure \ref{fig:GA}}, where part (a) shows a typical GA run in the form of the objective function being minimized plotted against the generation number (a maximum of 100 generations is run each time), and part (b) shows the PV efficiency plotted against E$_{gap}$ for compounds obtained from nearly 3000 different GA runs. It should be noted here that the GA models utilize the HSE-mf1 prediction for $\Delta$H and Expt-mf2 predictions for E$_{gap}$ and PCE, and mixing is allowed to happen in many different fractions, often simultaneously at A, B, and X sites. Every GA run requires 2 to 5 minutes on average for the 14-dimensional composition space with pre-trained RFR models for property prediction. We run GA models for four different subsets of the chemical space (D1, D2, D3, and D4), and show results for them in \textbf{Figure \ref{fig:GA}(b)}. These datasets have been explained in terms of their constituent chemical species in \textbf{Table \ref{table:GA}}; e.g., D2 represents Pb-free perovskites whereas D4 represents the much smaller space of (Cs-MA-FA)PbI$_3$ perovskites. \\

Since many different constraints are applied in the objective function, the best compounds obtained from a GA run may not be completely stoichiometrically accurate or meet every property target. As can be seen in \textbf{Figure \ref{fig:GA}(b)}, while all compounds essentially fulfill the E$_{gap}$ requirement (as well as the stability requirement, which is not plotted here), PCE is not always higher than 15\%. We find that across all the GA runs, there are a total of 1703 chemically meaningful ABX$_3$ alloys which satisfy conditions for all three properties; 484 of these compounds are cubic, 428 are tetragonal, 431 are orthorhombic, and 360 are hexagonal phase compounds. We also find that the total number of promising compounds increases when the chemical space becomes less complex---that is, a 5-dimensional space given by Cs-MA-FA-Pb-I leads to many more compounds with the targeted properties than the 14-dimensional space of K-Rb-Cs-MA-FA-Ca-Sr-Ba-Ge-Sn-Pb-I-Br-Cl. This is both a consequence of the difficulty of optimization in a high-dimensional space and the ease of finding attractive materials in the Cs-MA-FA-Pb-I space which are already among the best for stability and high efficiency. \textbf{Figure \ref{fig:GA}(b)} further shows that there are scores of Pb-free HaPs (D2) with high PV efficiencies, which opens up an important avenue towards eliminating Pb from solar cell perovskites. Many of the compounds are 14-dim, which means they are allowed to contain any species, while the remaining compounds are restricted to contain the most common species found in the experimental literature, namely Cs, MA, and FA at A-site, Pb, Sn, and Ge at the B-site, and Br/I at the X-site. \\

\begin{figure}[ht]
\includegraphics[width=1.0\linewidth]{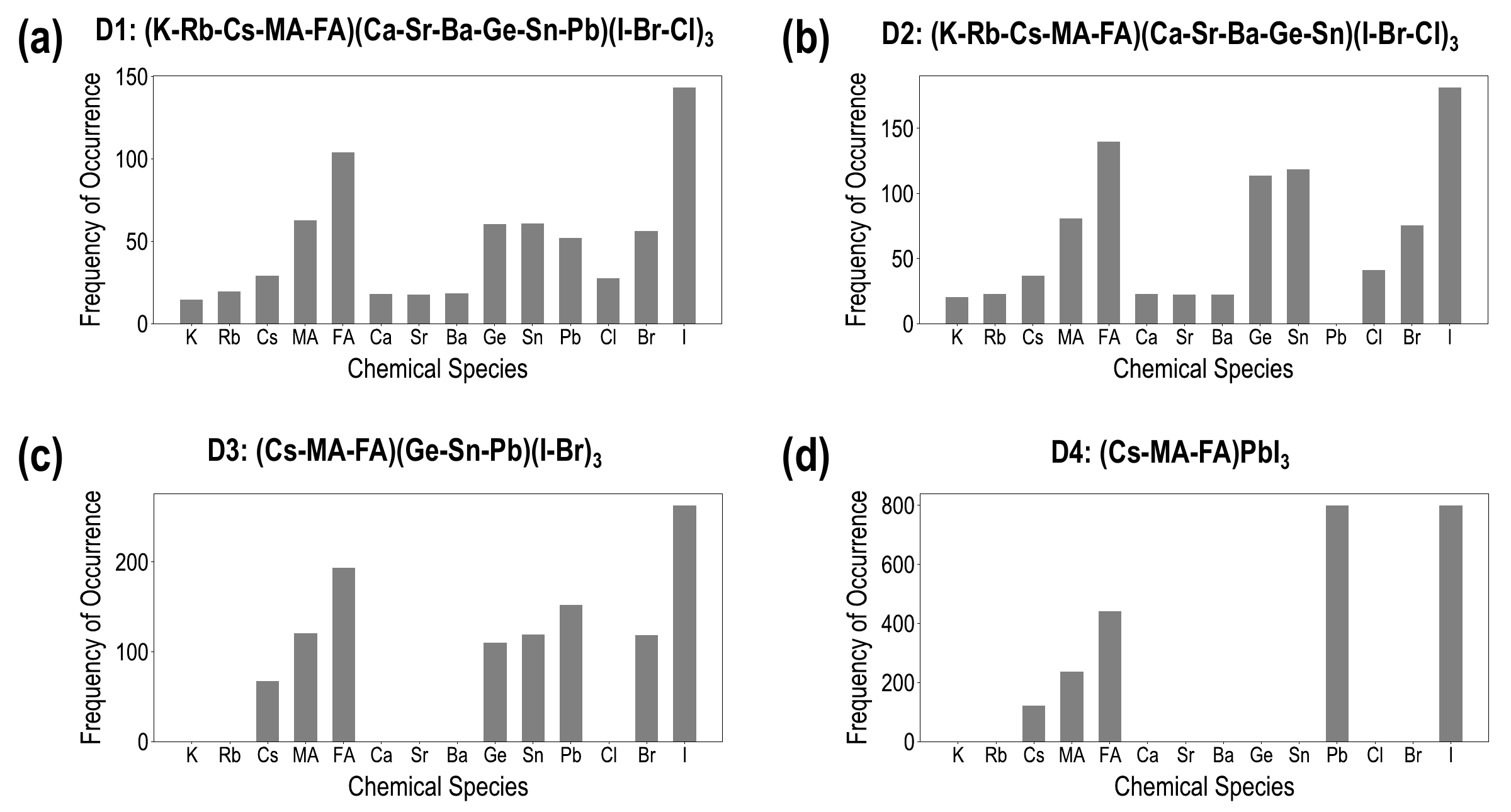}
\caption{\label{fig:GA_freq} Frequencies of occurrence of the 14 chemical species within the list of 1703 promising compounds obtained from GA, for subsets of the chemical space: (a) D1, (b) D2, (c) D3, and (d) D4.}
\end{figure}

An examination of the frequencies of occurrence in the GA-screened list of 1703 compounds is presented for the chemical spaces D1, D2, D3, and D4 in \textbf{Figure \ref{fig:GA_freq}}, with the results somewhat more interesting than the compounds screened from pure enumeration and prediction. In D1 and D2, we find FA and MA are the most prevalent at the A-site, followed by Cs, with Rb and K clearly preferred in smaller mixing fractions. Sn, Ge, and Pb are again the most common B-site cations, and Br makes a considerable appearance at X although I remains the king. In the total absence of Pb, Sn and Ge majorly dominate the B-site. \textbf{Figure \ref{fig:GA_freq}(a)} and \textbf{(b)} reveal that the best combination of properties could potentially be achieved using some combination of FA and MA at the A-site with small quantity of Cs, Sn and Ge at the B-site, and I with potentially small quantities of Br or Cl at the X-site. Further, the lower-dimensional chemical spaces D3 and D4 in \textbf{Figure \ref{fig:GA_freq}(c)} and \textbf{(d)} respectively show a preference for FA, Pb, and I, with Cs and Br being useful in smaller fractions, but Sn and Ge still holding their own.  \\

\begin{table}
    \centering
    \begin{tabular}{|c|c|c|c|c|c|}
\hline 
\textbf{Phase} & \textbf{All Screened Compounds} & \textbf{Pure} & \textbf{A-mixed} & \textbf{B-mixed} & \textbf{X-mixed} \\
\hline 
All & 3610 & 9 & 1364 & 2036 & 201 \\
Cubic & 1180 & 3 & 473 & 626 & 78 \\
Tetra & 856 & 2 & 290 & 510 & 54 \\
Ortho & 889 & 2 & 361 & 470 & 56 \\
Hex & 685 & 2 & 240 & 430 & 13 \\
\hline 
    \end{tabular}
    \caption{Number of ML-screened compounds with all properties in desirable ranges, namely $\Delta$H (HSE-mf1) < 0.2 eV, E$_{gap}$ (Expt-mf2) between 1 and 2 eV, and PCE (Expt-mf2) > 15\%. The screened compounds are divided based on their predicted phase and type of mixing.}
    \label{table:screened}
\end{table}

\begin{table}
    \centering
    \begin{tabular}{|c|c|c|c|c|c|c|}
\hline 
\textbf{Chemical Space} & \textbf{Description} & \textbf{Screened Compounds} & \textbf{Cubic} & \textbf{Tetragonal} & \textbf{Orthorhombic} & \textbf{Hexagonal} \\
\hline 
D1 & (K-Rb-Cs-MA-FA)(Ca-Sr-Ba-Ge-Sn-Pb)(I-Br-Cl)$_3$ & 226 & 80 & 51 & 58 & 37 \\
D2 & Pb-free, (K-Rb-Cs-MA-FA)(Ca-Sr-Ba-Ge-Sn)(I-Br-Cl)$_3$ & 297 & 90 & 70 & 71 & 66 \\
D3 & (Cs-MA-FA)(Ge-Sn-Pb)(I-Br)$_3$ & 381 & 114 & 107 & 102 & 58 \\
D4 & (Cs-MA-FA)PbI$_3$ & 799 & 200 & 200 & 200 & 199 \\
\hline 
    \end{tabular}
    \caption{The number of GA-screened compounds obtained for different chemical sub-spaces (D1, D2, D3, and D4), divided in terms of the perovskite phase.}
    \label{table:GA}
\end{table}

\begin{figure}[ht]
\includegraphics[width=1.0\linewidth]{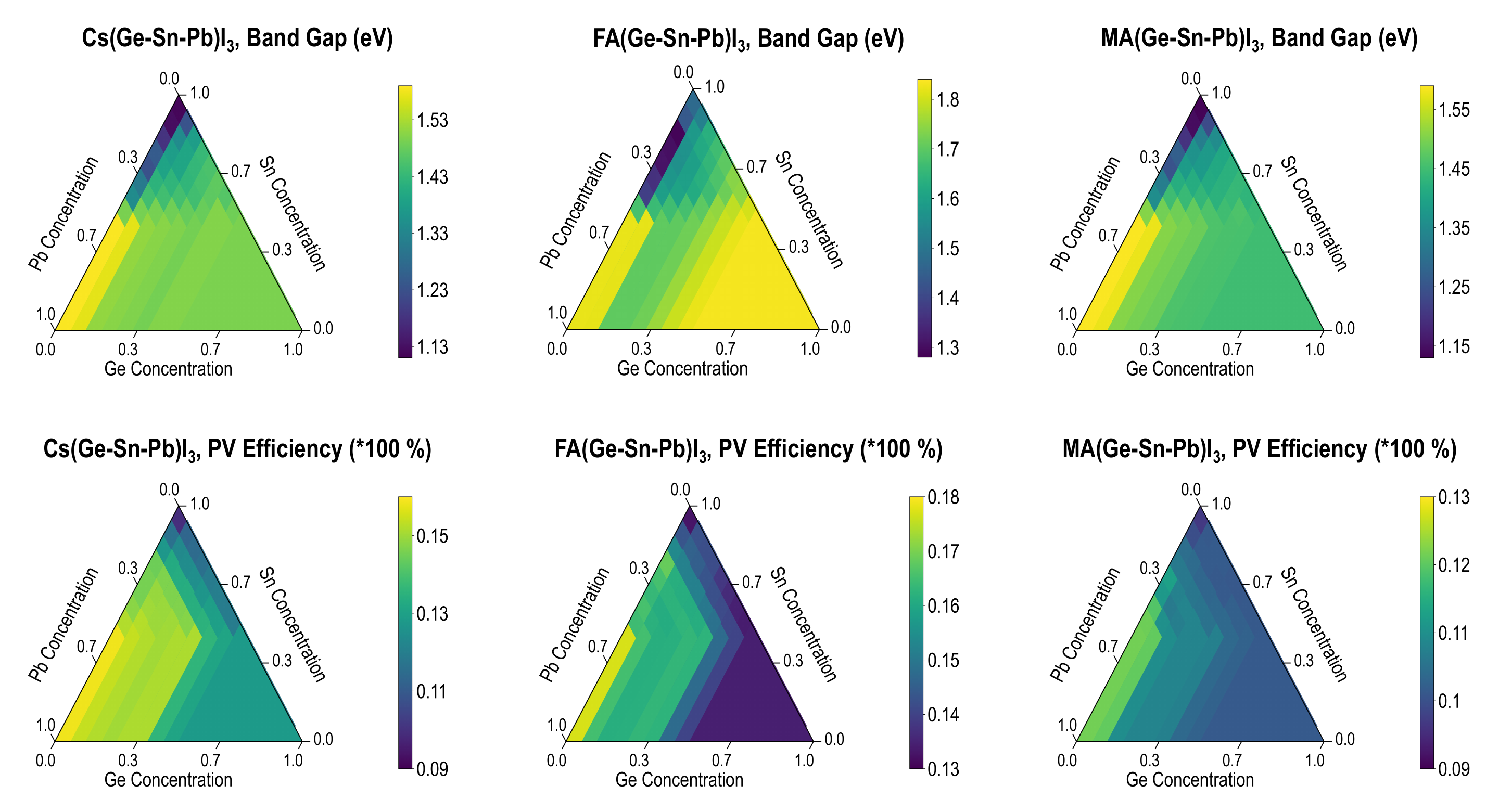}
\caption{\label{fig:ternary} Ternary property diagrams showing the band gap and PV efficiency from Expt-mf2 predictions, plotted for the chemical spaces Cs(Pb-Sn-Ge)I$_3$, FA(Pb-Sn-Ge)I$_3$, and MA(Pb-Sn-Ge)I$_3$.}
\end{figure}

\subsection*{\textbf{Visualizing Composition-Property Spaces using ML Predictions}}

Using ML predictions made across the set of 151,140 HaP compounds in different phases, it is now possible to visualize how the phase stability and computed properties vary in specific families of compounds, such as with increasing Sn concentration in Pb-Sn mixed compounds. Here, we pick a few interesting and important chemical spaces and visualize their property variations using ternary diagrams. \textbf{Figure \ref{fig:ternary}} shows the Expt-mf2 E$_{gap}$ and Expt-mf2 PV efficiency for the chemical spaces Cs(Ge-Sn-Pb)I$_3$, MA(Ge-Sn-Pb)I$_3$, and FA(Ge-Sn-Pb)I$_3$, plotted as ternary color maps with Ge, Sn, and Pb representing the three sides of the triangle. It can be seen that in the Cs(Ge-Sn-Pb)I$_3$ series, maximum E$_{gap}$ > 1.5 eV is shown by Pb-dominated compositions, minimum E$_{gap}$ $\sim$ 1.1 eV by Sn-dominated compositions, whereas Ge generally leads to more intermediate E$_{gap}$ values. MA(Ge-Sn-Pb)I$_3$ compounds show very similar E$_{gap}$ trends where in FA(Ge-Sn-Pb)I$_3$ compounds, a dominance of Pb or Ge leads to higher E$_{gap}$ values in the $\sim$ 1.8 eV range. The highest PV efficiencies are obtained by majority Pb compositions with some Sn and Ge added in all three series of compounds, leading to values > 12\% for MA(Ge-Sn-Pb)I$_3$, > 15\% for Cs(Ge-Sn-Pb)I$_3$, and $\sim$ 18\% for FA(Ge-Sn-Pb)I$_3$. It can also be seen that majority Ge compositions in FA(Ge-Sn-Pb)I$_3$ lead to much lower PV efficiencies around 13\%, so Pb-based compositions may still be crucial to achieve high efficiencies. \\

\textbf{Figures S7} to \textbf{S12} further show ternary plots for the phase stability (cubic vs tetra vs ortho vs hex), the HSE-mf1 $\Delta$H, and the Expt-mf2 E$_{gap}$ and PV efficiency, across the chemical spaces Cs(Ge-Sn-Pb)I$_3$, MA(Ge-Sn-Pb)I$_3$, FA(Ge-Sn-Pb)I$_3$, Cs(Ge-Sn-Pb)Br$_3$, MA(Ge-Sn-Pb)Br$_3$, and FA(Ge-Sn-Pb)Br$_3$. Such diagrams could be trivially plotted for any binary, ternary, or even higher-dimensional chemical spaces of interest, and can be navigated to determine the likely phase and bulk stability of any composition as well as their optoelectronic properties. It can be seen that the Cs-iodides generally prefer the tetragonal phase while the Cs-bromides prefer the orthorhombic phase. Nearly all compositions are stable w.r.t. decomposition in either series,  and the highest PV efficiencies $\sim$ 13\% for the bromides are achieved with Sn-dominance at the B-site. The MA-based iodides and bromides are nearly always tetragonal and show good bulk stability. Maximum efficiencies are obtained from mixed compounds with intermediate Pb fractions. Finally, the FA-based compounds are almost always stable in the hexagonal phase, except for a small area of cubic stability in the iodides. Sn-dominant FA-bromides provide a peak PV efficiency around 10\%, but the real pay-off comes from Pb-dominant iodides. The entire FA HaP space is highly stable against decomposition with $\Delta$H values < -1 eV p.f.u.

\section*{CONCLUSIONS}

In summary, several single-fidelity and multi-fidelity random forest regression models were trained for predicting the decomposition energy, band gap, and photovoltaic efficiency of halide perovskites at both DFT and experiment fidelity, resulting in the discovery of hundreds of promising new candidates via high-throughput screening and genetic algorithm. This work is based on an innovative approach of fusing experimental data from the literature with multiple types of DFT computed properties, and the unique representation of every material in terms of its composition and phase, well-known elemental or molecular properties of species at the cation and anion sites, and one-hot encoding fidelity information. All DFT data, optimized RFR models, and predictions made on hundreds of thousands of hypothetical compounds, are openly available to the community. GA enabled the efficient design of compositions beyond the limits of brute-force enumeration in certain mixing fractions, and provides an avenue for continuous discovery and improvement with infusion of more data and expanding the chemical space. Composition-property distributions are visualized in terms of ternary diagrams generated using ML predictions, revealing the most promising regions of B-site mixed (with Pb, Sn, and Ge) Cs/MA/FA iodides and bromides, and such exercises could easily be repeated for any chemical sub-spaces of interest. Limitations of the current work arise in terms of the experimental data only covering a small range of band gaps, any compound only represented as a single data point although it may posses dozens of polymorphs with varying properties, known issues with the DFT functionals, and the lack of consideration of other essential properties such as defect behavior and electron/hole mobilities. Each of these factors will be explored in future work; some of our ongoing work includes additional benchmarking of functionals for HaPs and the use of crystal graph-based neural networks for property prediction based on entire crystal structures as input. We anticipate that our datasets will serve many data mining and ML efforts, both within our group and from the community in general, and significant improvements will further be made in terms of data and ML approaches.

\section*{Conflicts of Interest}
There are no conflicts to declare.

\section*{Data Availability}
Tabulated data is included as spreadsheets in the supporting documents. All DFT data and ML models, including python scripts in Jupyter notebooks, can be found on Github: https://github.com/mannodiarun/perovs$\_$mfml$\_$ga.

\section*{Acknowledgements}
This work was performed at Purdue University, under startup account F.10023800.05.002 from the Materials Engineering department. This research used resources of the National Energy Research Scientific Computing Center (NERSC), the Laboratory Computing Resource Center (LCRC) at Argonne National Laboratory, and the Rosen Center for Advanced Computing (RCAC) clusters at Purdue. \\

\section*{References}

\nocite{*}
\bibliography{aipsamp}% Produces the bibliography via BibTeX.

\clearpage
\newpage
\pagenumbering{gobble}
\thispagestyle{empty} 

%\onecolumngrid
%\onecolumn

\setcounter{figure}{0}   
\setcounter{table}{0} 
\renewcommand{\thetable}{S\Roman{table}} 
\renewcommand\thefigure{S\arabic{figure}}

\begin{center}
\vspace{0.5cm}
\Large
\textbf{Supplemental material to "Discovering Novel Halide Perovskite Alloys using Multi-Fidelity Machine Learning and Genetic Algorithm"\\}
\vspace{0.5cm}
\large
Jiaqi Yang,\textsuperscript{1}, Panayotis Manganaris,\textsuperscript{1} and Arun Mannodi-Kanakkithodi,\textsuperscript{1}\\
\vspace{0.3cm}

\normalsize
\textsuperscript{1}\textit{School of Materials Engineering, Purdue University, West Lafayette, Indiana 47907, USA}\\
\end{center}

%\footnote{
%\textsuperscript{a}amannodi@purdue.edu}

\vspace{1cm}

\begin{table}[!ht]
    \centering
    \begin{tabular}{|c|c|}
    \hline
        \textbf{Descriptor Label} & \textbf{Descriptor Meaning} \\ \hline
        K\_frac & K fraction over A sites \\ \hline
        Rb\_frac & Rb fraction over A sites \\ \hline
        Cs\_frac & Cs fraction over A sites \\ \hline
        MA\_frac & MA fraction over A sites \\ \hline
        FA\_frac & FA fraction over A sites \\ \hline
        Ca\_frac & Ca fraction over B sites \\ \hline
        Sr\_frac & Sr fraction over B sites \\ \hline
        Ba\_frac & Ba fraction over B sites \\ \hline
        Ge\_frac & Ge fraction over B sites \\ \hline
        Sn\_frac & Sn fraction over B sites \\ \hline
        Pb\_frac & Pb fraction over B sites \\ \hline
        Cl\_frac & Cl fraction over X sites \\ \hline
        Br\_frac & Br fraction over X sites \\ \hline
        I\_frac & I fraction over X sites \\ \hline
        Cubic & 1 if cubic, 0 if not \\ \hline
        Tetra & 1 if tetragonal, 0 if not \\ \hline
        Ortho & 1 if orthorhombic, 0 if not \\ \hline
        Hex & 1 if hexagonal, 0 if not \\ \hline
        PBE & 1 if data is from PBE, 0 if not \\ \hline
        HSE & 1 if data is from HSE, 0 if not \\ \hline
        Expt & 1 if data is from Expt, 0 if not \\ \hline
        A\_ion\_rad & Ion radius of A site elements \\ \hline
        A\_BP & Boiling point of A site elements \\ \hline
        A\_MP & Melting point of A site elements \\ \hline
        A\_dens & Density of A site elements \\ \hline
        A\_at\_wt & Atomic weight of A site elements \\ \hline
        A\_EA & Electron affinity of A site elements \\ \hline
        A\_IE & Ionization energy of A site elements \\ \hline
        A\_hof & Heat of formation of A site elements \\ \hline
        A\_hov & Heat of vaporization of A site elements \\ \hline
        A\_En & Electronegativity of A site elements \\ \hline
        A\_at\_num & Atomic number of A site elements \\ \hline
        A\_period & Period number of of A site elements \\ \hline
        B\_ion\_rad & Ion radius of B site elements \\ \hline
        B\_BP & Boiling point of B site elements \\ \hline
        B\_MP & Melting point of B site elements \\ \hline
        B\_dens & Density of B site elements \\ \hline
        B\_at\_wt & Atomic weight of B site elements \\ \hline
        B\_EA & Electron affinity of B site elements \\ \hline
        B\_IE & Ionization energy of B site elements \\ \hline
        B\_hof & Heat of formation of B site elements \\ \hline
        B\_hov & Heat of vaporization of B site elements \\ \hline
        B\_En & Electronegativity of B site elements \\ \hline
        B\_at\_num & Atomic number of B site elements \\ \hline
        B\_period & Period number of of B site elements \\ \hline
        X\_ion\_rad & Ion radius of X site elements \\ \hline
        X\_BP & Boiling point of X site elements \\ \hline
        X\_MP & Melting point of X site elements \\ \hline
        X\_dens & Density of X site elements \\ \hline
        X\_at\_wt & Atomic weight of X site elements \\ \hline
        X\_EA & Electron affinity of X site elements \\ \hline
        X\_IE & Ionization energy of X site elements \\ \hline
        X\_hof & Heat of formation of X site elements \\ \hline
        X\_hov & Heat of vaporization of X site elements \\ \hline
        X\_En & Electronegativity of X site elements \\ \hline
        X\_at\_num & Atomic number of X site elements \\ \hline
        X\_period & Period number of of X site elements \\ \hline
    \end{tabular}
    \caption{\label{table:SI_list_desc} Explanation of every descriptor dimension used in this work, enabling the development of RFR models and subsequent predictions on new materials.}
\end{table}

\begin{figure*}[ht]
\centering
\includegraphics[width=0.8\linewidth]{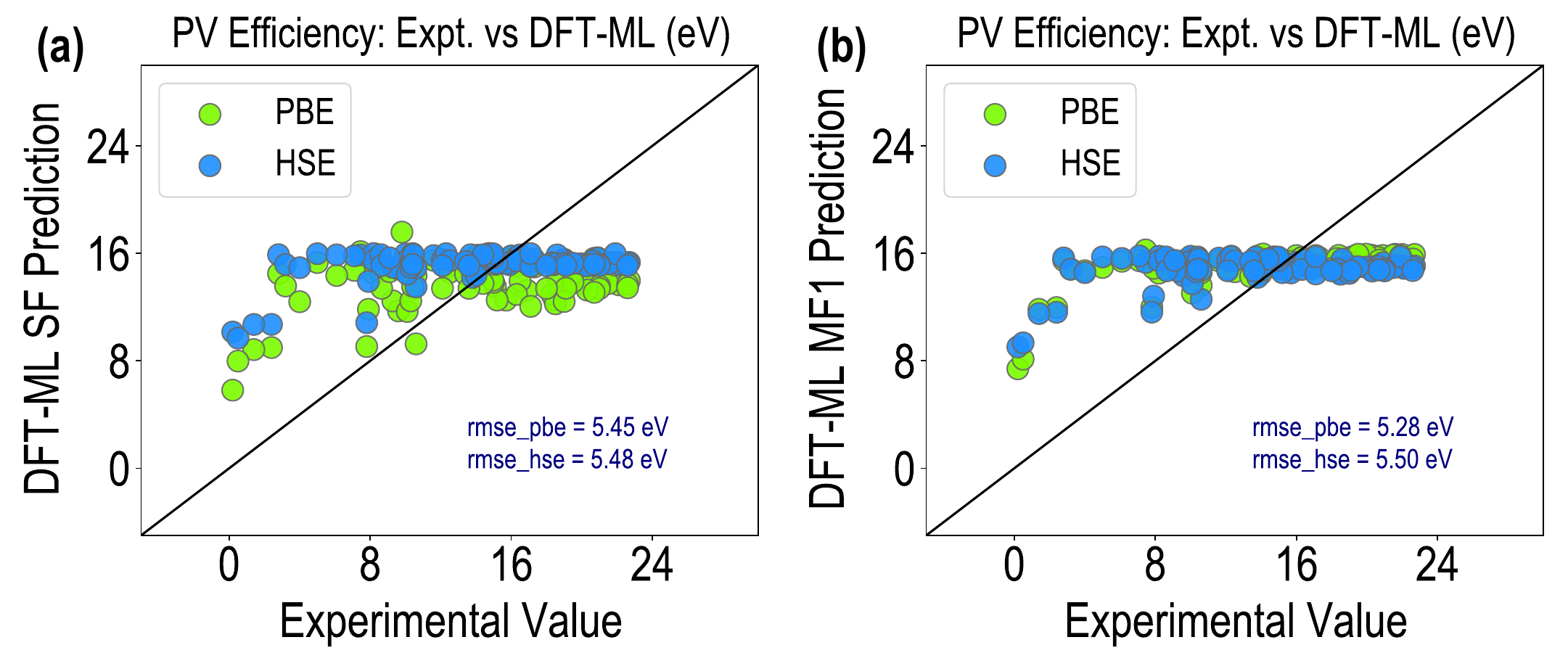}
\caption{\label{Fig:SI_expt_dft_slme}
DFT-ML SLME plotted against experimental PCE for (a) single-fidelity RFR model, and (b) multi-fidelity RFR model.}
\end{figure*}

\begin{figure*}[ht]
\centering
\includegraphics[width=\linewidth]{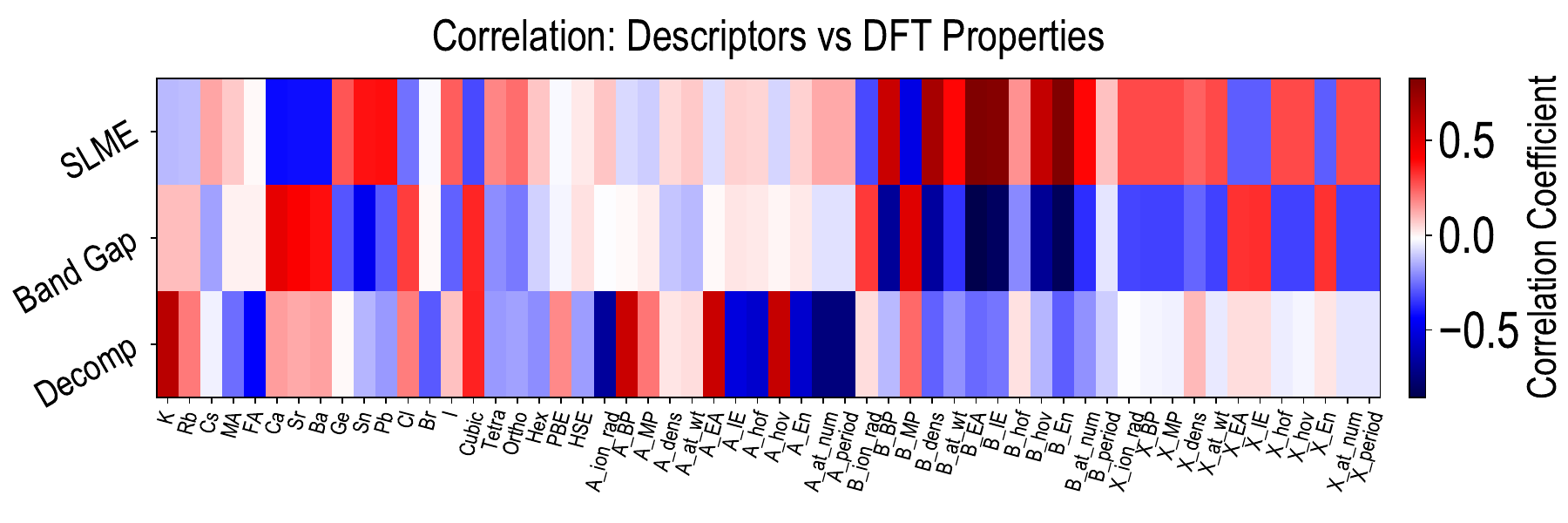}
\caption{\label{Fig:SI_corr}
Pearson coefficients of linear correlation between the 56-dimensional descriptors (for the PBE+HSE dataset) and the three properties computed from PBE and HSE.}
\end{figure*}

\begin{figure*}[ht]
\centering
\includegraphics[width=0.8\linewidth]{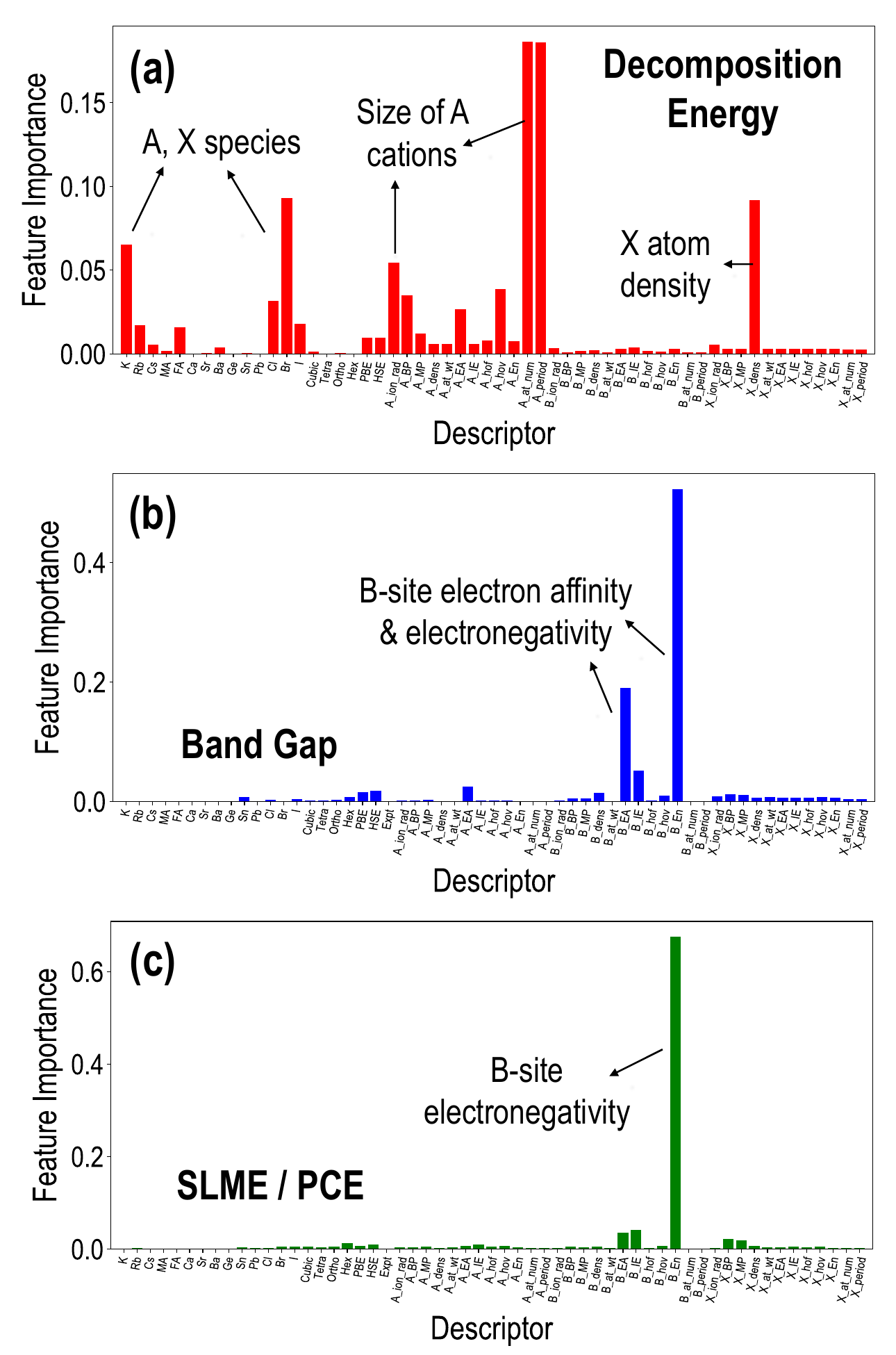}
\caption{\label{Fig:SI_rfr_imp}
Importance values of each feature from multi-fidelity random forest regression trained on the combined PBE, HSE, and Expt dataset, for (a) decomposition energy, (b) band gap, and (c) PV efficiency.}
\end{figure*}

\begin{figure*}[ht]
\centering
\includegraphics[width=0.8\linewidth]{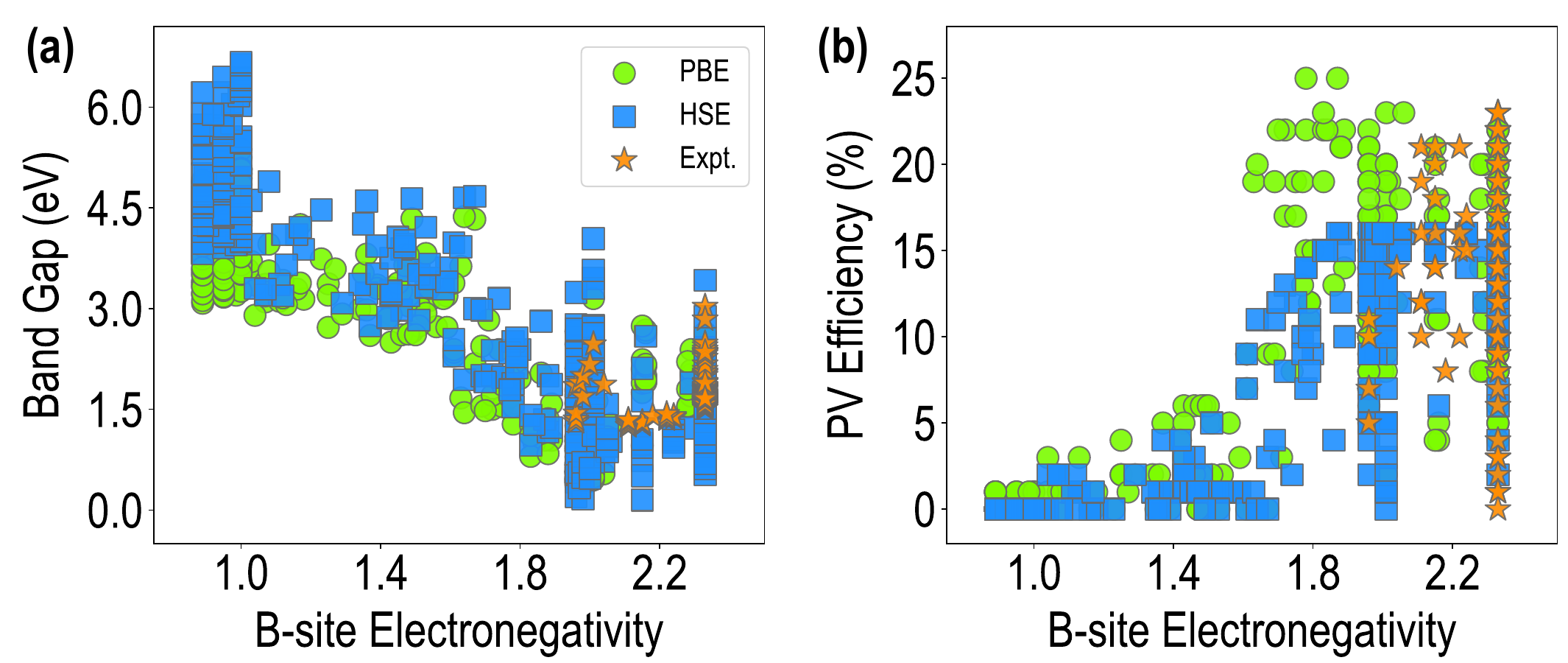}
\caption{\label{Fig:SI_EN}
PBE, HSE, and expt. properties plotted against B-site electronegativity: (a) band gap, and (c) PV efficiency.}
\end{figure*}

\begin{figure*}[ht]
\centering
\includegraphics[width=0.8\linewidth]{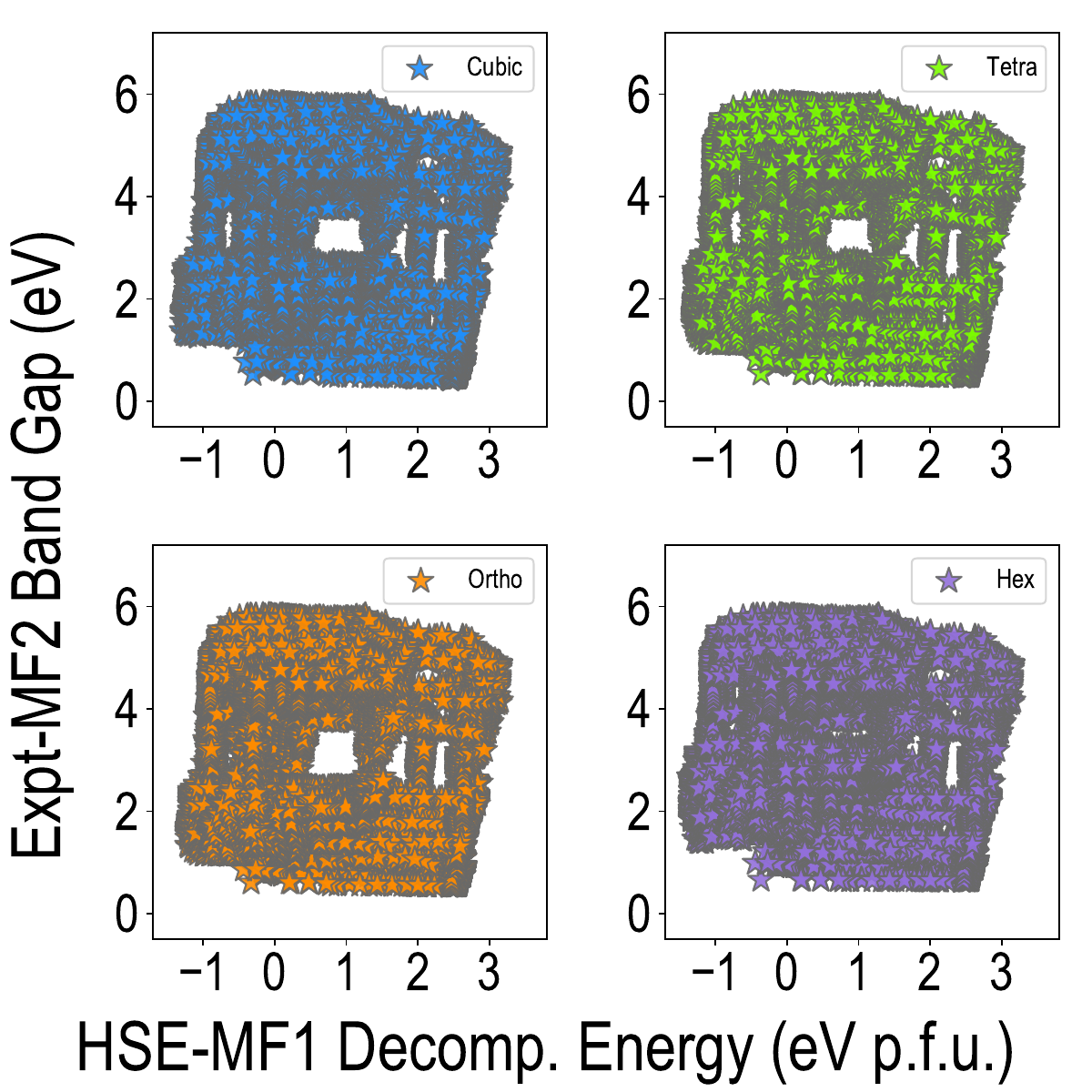}
\caption{\label{Fig:SI_out_1}
Visualization of the ML-predicted properties across the expanded chemical space of $\sim$ 150,000 compounds, in terms of band gap (Expt-mf2) vs decomposition energy (HSE-mf1) plots.}
\end{figure*}

\begin{figure*}[ht]
\centering
\includegraphics[width=0.8\linewidth]{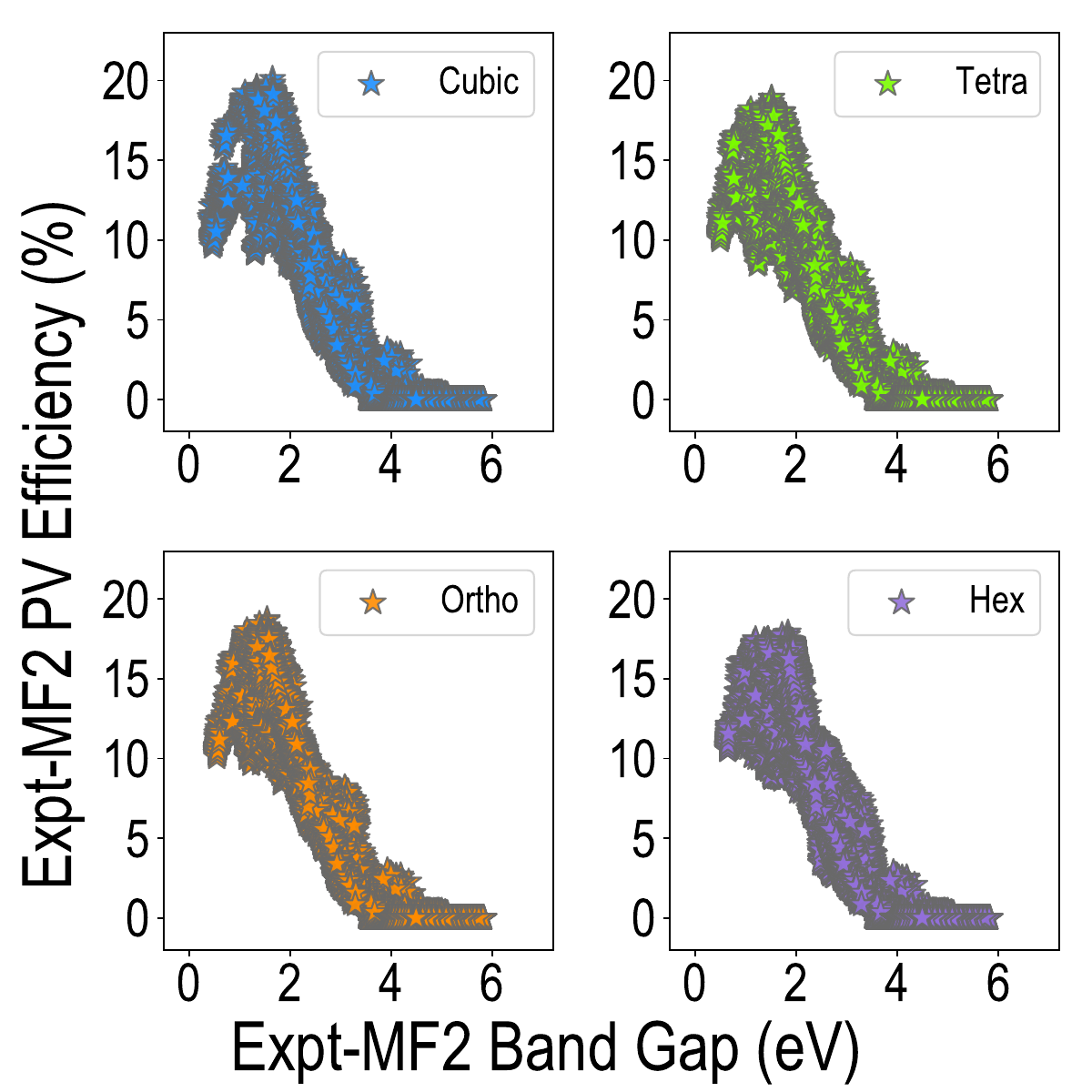}
\caption{\label{Fig:SI_out_2}
Visualization of the ML-predicted properties across the expanded chemical space of $\sim$ 150,000 compounds, in terms of PV efficiency (Expt-mf2) vs band gap (Expt-mf2) plots.}
\end{figure*}

\begin{figure}[ht]
\includegraphics[width=1.0\linewidth]{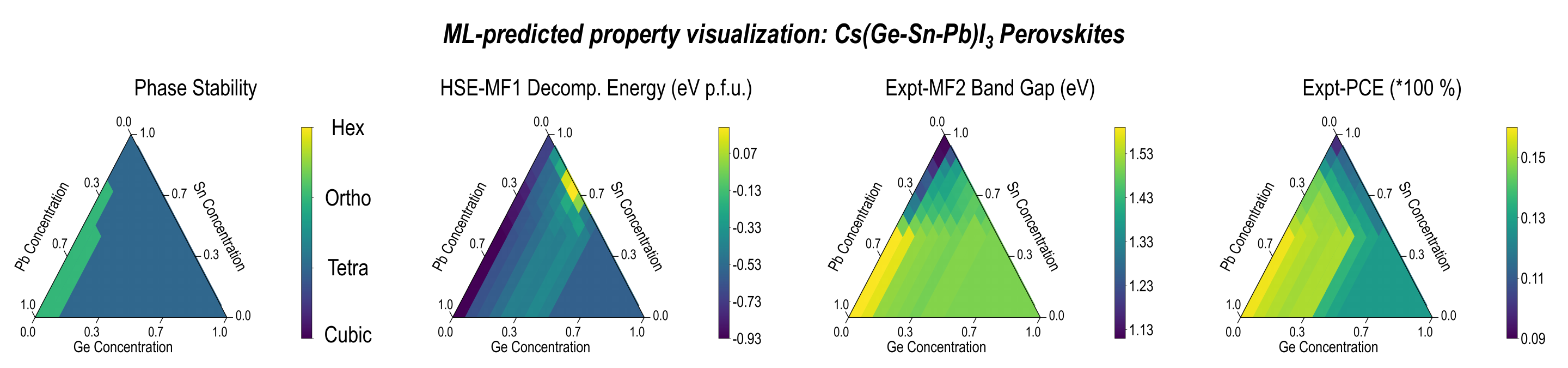}
\caption{\label{fig:tern_Cs_I} Ternary property diagrams showing the predicted phase stability, decomposition energy, band gap, and PV efficiency for Cs(Ge-Sn-Pb)I$_3$ perovskites.}
\end{figure}

\begin{figure}[ht]
\includegraphics[width=1.0\linewidth]{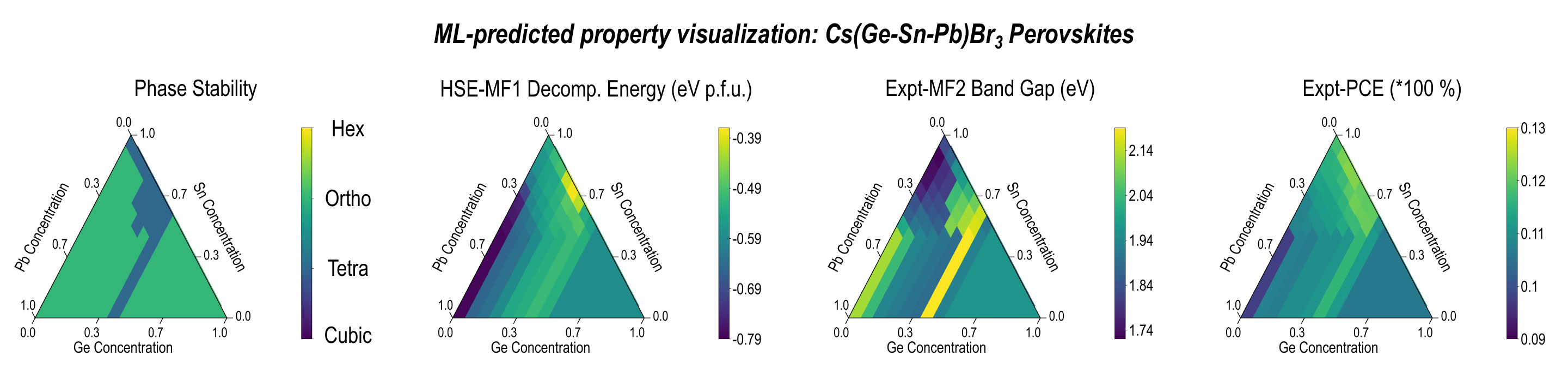}
\caption{\label{fig:tern_Cs_Br} Ternary property diagrams showing the predicted phase stability, decomposition energy, band gap, and PV efficiency for Cs(Ge-Sn-Pb)Br$_3$ perovskites.}
\end{figure}

\begin{figure}[ht]
\includegraphics[width=1.0\linewidth]{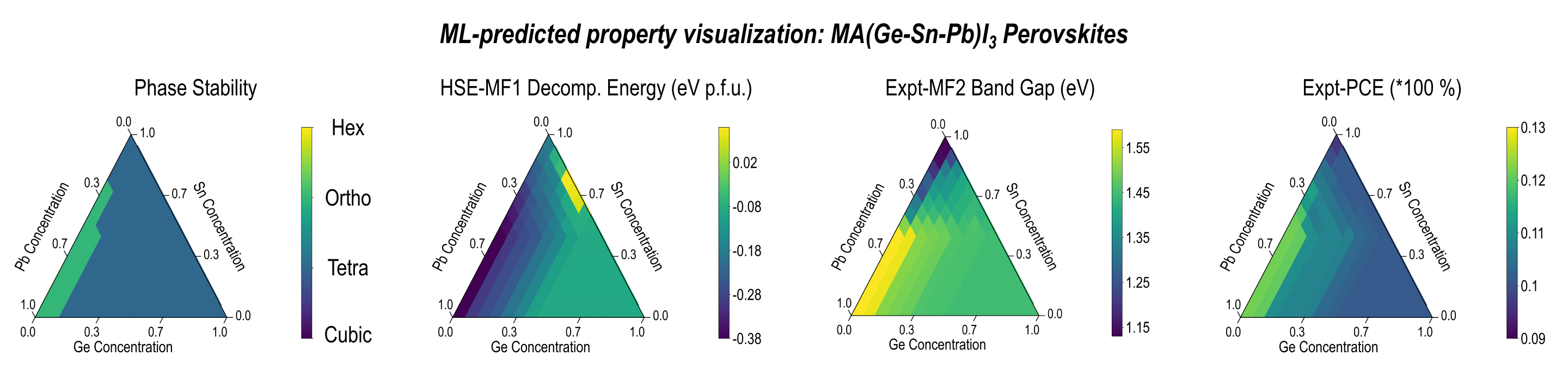}
\caption{\label{fig:tern_MA_I} Ternary property diagrams showing the predicted phase stability, decomposition energy, band gap, and PV efficiency for MA(Ge-Sn-Pb)I$_3$ perovskites.}
\end{figure}

\begin{figure}[ht]
\includegraphics[width=1.0\linewidth]{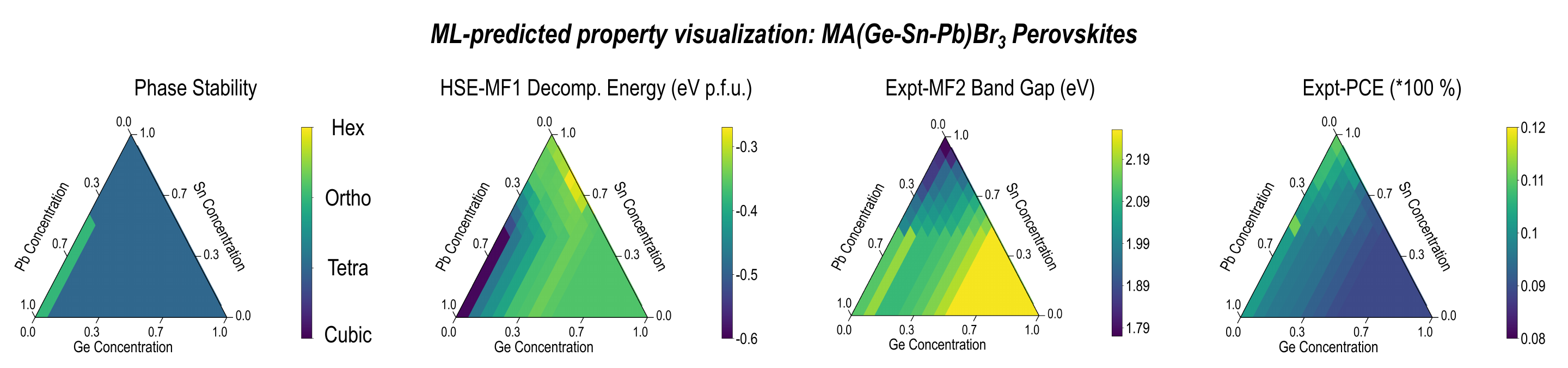}
\caption{\label{fig:tern_MA_Br} Ternary property diagrams showing the predicted phase stability, decomposition energy, band gap, and PV efficiency for MA(Ge-Sn-Pb)Br$_3$ perovskites.}
\end{figure}

\begin{figure}[ht]
\includegraphics[width=1.0\linewidth]{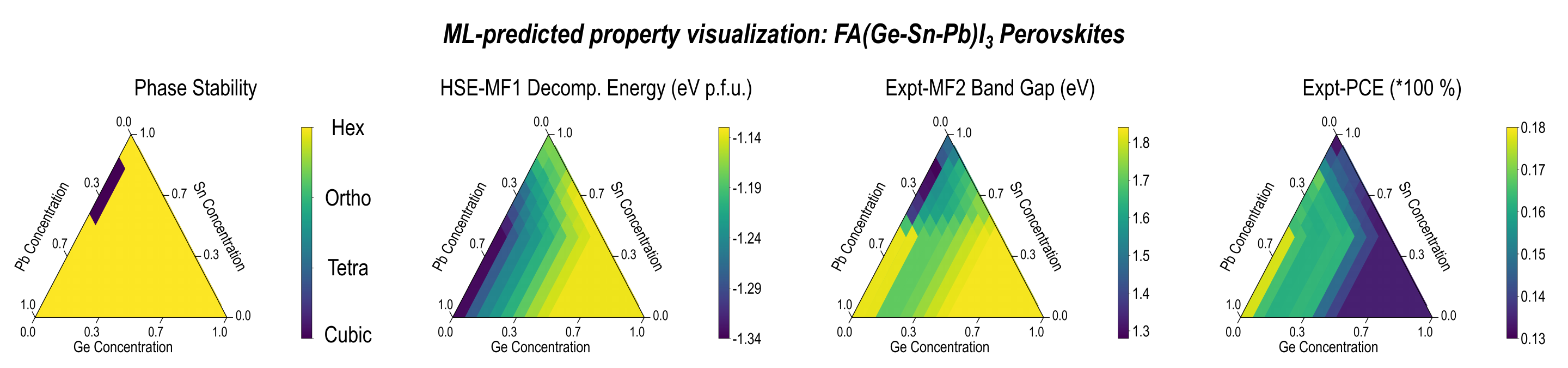}
\caption{\label{fig:tern_FA_I} Ternary property diagrams showing the predicted phase stability, decomposition energy, band gap, and PV efficiency for FA(Ge-Sn-Pb)I$_3$ perovskites.}
\end{figure}

\begin{figure}[ht]
\includegraphics[width=1.0\linewidth]{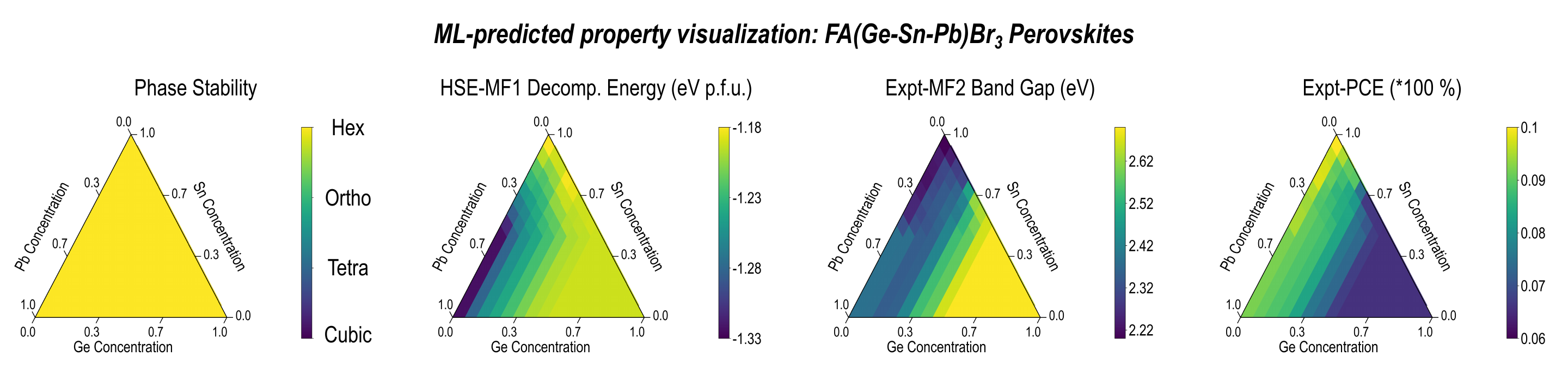}
\caption{\label{fig:tern_FA_Br} Ternary property diagrams showing the predicted phase stability, decomposition energy, band gap, and PV efficiency for FA(Ge-Sn-Pb)Br$_3$ perovskites.}
\end{figure}

\thispagestyle{empty}

\end{document}